\documentclass[aps,prd,amssymb,twocolumn,superscriptaddress,floatfix,nofootinbib,10pt]{revtex4-1}
\usepackage{graphicx}
\usepackage{dcolumn}
\usepackage{hyperref}
\usepackage{ulem}
\usepackage{amsmath}
\usepackage{amsfonts}
\usepackage{booktabs}
\usepackage{amssymb}
\usepackage[usenames]{color}
\usepackage{lineno}
\usepackage{bm}
\usepackage{mathptmx}
\usepackage{multirow}
\usepackage{soul}

\begin{document}

\preprint{APS/123-QED}

\title{Femtoscopic study of the $S=-1$ meson-baryon interaction: $K^-p$, $\pi^-\Lambda$ and $K^+\Xi^-$ correlations}

\author{P. Encarnación}
	\email{Pablo.Encarnacion@ific.uv.es}
	\affiliation{Departament de F\'{i}sica Qu\`antica i Astrof\'{i}sica and Institut de Ci\`encies del Cosmos (ICCUB), Facultat de F\'{i}sica, Universitat de Barcelona, Barcelona, Spain}
     \affiliation{Departamento de F\'{i}sica Teórica and IFIC, Centro Mixto Universidad de Valencia-CSIC, Institutos de Investigaci\'{o}n de Paterna, Aptdo. 22085, E-46071 Valencia, Spain}
 
\author{A. Feijoo}
 \thanks{Corresponding author}
	\email{edfeijoo@ific.uv.es}
	\affiliation{Physik Department E62, Technische Universit\"at M\"unchen, Garching, Germany, EU}
    \affiliation{Instituto de F\'{i}sica Corpuscular, Centro Mixto Universidad de Valencia-CSIC, Institutos de Investigaci\'{o}n de Paterna, Aptdo. 22085, E-46071 Valencia, Spain}

\author{ V. Mantovani Sarti}
 \email{valentina.mantovani-sarti@tum.de}
	\affiliation{Physik Department E62, Technische Universit\"at M\"unchen, Garching, Germany, EU}

 \author{A. Ramos}
	\email{ramos@fqa.ub.edu}
	\affiliation{Departament de F\'{i}sica Qu\`antica i Astrof\'{i}sica and Institut de Ci\`encies del Cosmos (ICCUB), Facultat de F\'{i}sica, Universitat de Barcelona, Barcelona, Spain}
 
\date{\today}

\begin{abstract}
We study the femtoscopic correlation functions of meson-baryon pairs in the strangeness $S=-1$ sector, employing unitarized s-wave scattering amplitudes derived from the chiral Lagrangian up to next-to-leading order.  For the first time, we deliver predictions on the $\pi^-\Lambda$ and $K^+\Xi^-$ correlation functions which are feasible to be measured at the Large Hadron Collider. We also demonstrate that the employed model is perfectly capable of reproducing the $K^-p$ correlation function data measured by the same collaboration, without the need to modify the coupling strength to the $\bar{K}^0n$ channel, as has been recently suggested. In all cases, the effects of the source size on the correlation are tested. In addition, we present detailed analysis of the different coupled-channel contributions, together with the quantification of the relative relevance of the different terms in the interaction. These calculations require the knowledge of the so-called production weights, for which we present two novel methods to compute them. 
\end{abstract}

\maketitle


\section{Introduction}
\label{sec:introduction}

The scattering between kaons and nucleons has drawn the attention of the theoretical community in the last few decades \cite{dalitz,thomas,fink,rubin}. The attractive character of the $\bar{K}N$ interaction and the presence of a large number of resonant states around the $\bar{K}N$ threshold offer a perfect testing ground for the chiral unitary approach. Indeed, the Unitaritzed Chiral Perturbation Theory framework (UChPT) has been shown to be a powerful tool to treat the low-energy meson-baryon interaction in the $S=-1$ sector. The success of this non-perturbative scheme lies in the ability to reproduce the experimental data and the dynamic generation of bound states and resonances, as already proved in the early stages of this approach \cite{KSW,KWW,OR,OM,LK,BMW,Garcia-Recio:2002yxy,BFMS,BNW}. Among the generated states, the most outstanding one is the $\Lambda(1405)$ resonance for its intricate nature, whose most plausible interpretation comes in terms of a double-pole contribution arising from coupled-channel meson-baryon re-scattering \cite{OM,2pole,PRL}. To illustrate the controversy around this state, it suffices to mention that although the dynamical origin of $\Lambda(1405)$ was predicted in the late 1950s \cite{L1405}, this interpretation has found its way into the PDG compilation \cite{ParticleDataGroup:2024cfk} only recently.

Continuing chronologically, one of the main messages appearing in Ref.~\cite{BMN} was the need for additional subthreshold information on the antikaon-nucleon dynamics to correctly locate the two poles associated to the $\Lambda(1405)$ state. In order to amend the lack of constraints, the experimental machinery was set in motion, and several groups carried out new measurements that certainly shed some light on this topic. The $\pi\Sigma$ mass distributions from $pp$ scattering experiments were provided by the COSY and HADES collaborations, \cite{Zychor:2007gf} and \cite{HADES:2012csk} respectively. Additional information came from JLAB, where photo-production differential cross sections for the $\Sigma(1385)$, $\Lambda(1405)$, and $\Lambda(1520)$ in the reactions $\gamma + p \to K^+ + Y^*$ were provided using the CLAS detector. Furthermore, the CLAS collaboration also reported on a direct determination of the expected spin-parity $J^\pi=1/2^-$ of the $\Lambda(1405)$ \cite{CLAS:2014tbc}. However, the SIDDHARTA collaboration \cite{SIDD} performed the most striking measurement to constrain the theoretical models, which consists of the precise determination of the energy shift and width (around $20$\%) of the $1s$ state in kaonic hydrogen. In this way, the tension between the DEAR \cite{DEAR:2005fdl,DEAR:2005ueb} and KEK \cite{Iwasaki:1997wf} measurements, with almost a factor two of relative uncertainty, could be resolved. The availability of these experimental data again boosted the theoretical community \cite{IHW,HJ_rev,Cieply:2011nq,GO,Mizutani:2012gy,Roca:2013av,Roca:2013cca,Mai:2014xna}, and the models were revisited, in some cases extended to higher orders and energies, with the aim of describing the observables within the new experimental uncertainties. It should be noted that, if the models are limited to accommodate the two-body cross sections of $K^- p$ scattering into $\pi\Sigma, \bar{K}N,\pi\Lambda$ states (the classical channels) or the experimental photo-production data on $\gamma p \to K^+ \pi \Sigma$ reactions, the dominant contribution to reach a good agreement with the experimental data is the contact Weinberg-Tomozawa (WT) one. In other words, the incorporation of other ($O(p)$) corrections, i. e. the direct and crossed Born terms, as well as the next-to-leading order (NLO) terms plays merely a fine-tuning role. In fact, the model in Ref.~\cite{GO} was constrained by a larger set of data that included, apart from the classical cross sections and the scattering length extracted from the SIDDHARTA outputs, the scattering data from $K^- p \to \eta \Lambda , \pi^0 \pi^0 \Sigma^0 $ and data from two event distributions ($K^- p \to \Sigma^+(1660) \pi^- \to \Sigma^+ \pi^- \pi^+ \pi^-$ and $ K^- p \to \pi^0 \pi^0 \Sigma^0$). From there, one immediately realizes that the inclusion of NLO contribution improves remarkably the reproduction of the $K^- p \to \eta \Lambda$ scattering data.

One of the challenges to be faced when incorporating the NLO terms of the chiral lagrangian is the determination of the corresponding low energy constants (LECs), which are not established by the underlying theory and should be obtained through fitting procedures to experimental data sensitive to these higher order corrections. The $K^- p \to K \Xi$ reactions are an example of such kind of processes, since they do not proceed via the WT term and the rescattering terms due to the coupled channels taken in the Bethe-Salpeter (BS) equation are not sufficient to reproduce the experimental cross section. In the series of papers \cite{Feijoo:2015yja,Ramos:2016odk}, the authors not only demonstrate the sensitivity of the $K \Xi$ channels to the NLO terms but also obtained results that revealed the particular relevance of the u- and s-diagrams. In similar spirit, the global study \cite{Lu:2022hwm} of the $S=0,-1,+1$ meson-baryon sectors based on covariant baryon chiral perturbation theory up to next-to-next-to-leading order put tighter constraints on the amplitudes and on the resonances generated. A step further was given in \cite{Feijoo:2018den}, where, motivated by the findings of the former studies and aiming at more reliable values of the NLO coefficients, the $K^- p \to \eta \Lambda, \eta \Sigma^0$ reactions were incorporated in the fits thereby having information from all possible channels of the $S=-1$ sector. The model obtained (BCN model) is able to reproduce all the available low-energy scattering data from $K^-p\to (S=-1)$~pseudoscalar-baryon processes with a very reasonable agreement, as well as all the $\bar{K}N$ threshold observables typically employed in these studies (branching ratios and the scattering length extracted from the SIDDHARTA measurements \cite{SIDD}). The BCN model also does a good job in reproducing within errors the strength of the $K^- n \to \pi^- \Lambda$ amplitude ($30$~MeV below $\bar{K}N$ threshold), which was extrapolated from $K^-$ absorption processes on $^{4}$He \cite{Piscicchia:2018rez}. Aiming at testing the validity of the BCN model at higher energies, a prediction of the $K^0_L p \to K^+ \Xi^0$ cross section, proposed for its measurement at Jlab \cite{Amaryan:2016ufk}, was also given. Moreover, when implementing in-medium corrections including one- and two-nucleon absorption channels, the BCN model reproduces 
satisfactorily the antikaon absorption rates in $^{12}$C \cite{Hrtankova:2019jky} and improves significantly the description of kaonic atom data \cite{Obertova:2022wpw}.

Unfortunately, despite all previous experimental and theoretical advances, if one turns the attention to the amplitude behavior below the $\bar{K}N$ threshold and compares what comes out employing different chirally motivated benchmark models, notable discrepancies can be appreciated. This is clearly illustrated in Fig.~1 of Ref.~\cite{Cieply:2020ftt}, which is particularized for $K^-p$ and $K^-n$ elastic processes and shows how, due to the experimental constraints, the models converge from the $\bar{K}N$ threshold on while diverging substantially below it. As a direct consequence of the ill-determined $\bar{K}N$ subthreshold amplitudes, the uncertainty associated to the position of the lower mass $\Lambda(1405)$ pole becomes very large. More precisely, as can be seen in Fig.~1 and Fig.~6 of Refs.~\cite{Cieply:2016jby,Feijoo:2018den} respectively, the different models produce very scattered locations in the complex plane for this pole, while the second pole seems to be very well pinned down since all models provide coordinates gathered around $1420$~MeV and with a width of approximately $40$~MeV. An extensive overview on this topic can be found in \cite{Mai:2020ltx}.

In the last years many efforts have been made to improve this situation. This is evidenced by the numerous experiments underway or planned, as well as the theoretical works devoted to this topic that can be found in the literature recently.

With respect to photoproduction experiments, it is worth mentioning the latest GlueX preliminary analysis of the $\pi^0\Sigma^0$ invariant mass from the $\gamma p \to K^+\pi^0\Sigma^0$ process \cite{Wickramaarachchi:2022mhi}, which is in agreement with the previous experimental and theoretical evidences about the double pole structure of the $\Lambda(1405)$. This work was preceded by the theoretical study (and the subsequent extension of the formalism) of $\gamma p \to K^+\pi \Sigma$ photo-production mechanism, in Refs.~\cite{Bruns:2022sio,Bruns:2024ivd}, following an approach that incorporates constraints from unitarity, gauge invariance and chiral perturbation theory.

The next important experimental output will come from the SIDDHARTA-2 high precision measurement of the X-ray of the $2p$ to $1s$ transition in kaonic deuterium. The data acquisition campaign is presently ongoing \cite{Sgaramella:2024qdh}. The combination of this challenging measurement with that of the kaonic hydrogen measurement \cite{SIDD} will provide the isospin-dependent antikaon-nucleon scattering lengths at threshold, which represents a milestone in the exploration of QCD at low energies with strangeness.

Another source of knowledge for the $\bar{K}N$ interaction below threshold comes from the antineutrino induced $\Lambda(1405)$ given the relevant role of the strong interaction in the rescattering. This reaction can go via the process $\bar{\nu}p \to l^+\phi B$ ($\phi B$ being a meson-baryon pair) , which was theoretically studied in Ref.~\cite{Ren:2015bsa}. On the experimental side, this reaction is one of the possible outputs of the MicroBooNE collaboration, where the role of $\Lambda$, $\Sigma$ and related hyperon production is currently under analysis. In addition, it is expected that the SBND detector at Fermilab will be able to measure such processes with huge statistics.

The $K^-d\to p\Sigma^-$ reaction was suggested in \cite{Feijoo:2021jtr} as an alternative window to the $\bar{K}N$ subthreshold amplitudes. This reaction should take place by means of two triangle topologies which develop a triangle singularity. The authors compute the total cross-section of the process for different $\bar{K}N$ amplitudes calculated within the UChPT approach. The difference among the different total cross sections is large enough to claim that a future comparison with the experiment could play an important role to discern which models are the most suitable to describe the physics below the $\bar{K}N$ threshold, with clear implications to pin down the lower mass pole of the $\Lambda(1405)$.

In this respect, the Lattice community also makes its contribution providing the simulation on the $I=0$ coupled-channel scattering amplitudes of $\pi\Sigma-\bar{K} N$ \cite{BaryonScatteringBaSc:2023ori,BaryonScatteringBaSc:2023zvt}. The two-channel K-matrix fitted to the lattice QCD data supported the two-pole picture in agreement with UChPT. Slightly after, a theoretical finite volume analysis based on chiral lagrangians \cite{Zhuang:2024udv} reached a remarkable consistency between the chiral unitary approach predictions for the two-pole structure, the recent LQCD scattering data and the available experimental cross sections. In \cite{Ren:2024frr}, another theoretical analysis within the renormalizable framework of covariant chiral effective field theory obtained both $\Lambda(1405)$ poles compatible with those provided by the BaSc Collaboration \cite{BaryonScatteringBaSc:2023ori,BaryonScatteringBaSc:2023zvt}.

The authors of \cite{Nieves:2024dcz} presented a detailed discussion on the lowest-lying ${\frac{1}{2}}^-$ and ${\frac{3}{2}}^-$ $\Lambda_{Q}$ resonances  ($Q=s,c,b$), paying special attention to the interplay between the conventional quark model (CQM) and chiral baryon-meson degrees of freedom, which are coupled using a unitarized scheme consistent with leading order (LO) heavy quark symmetries. The main conclusion is that the two-pole pattern in the strange sector is a consequence of the decisive role of the $\bar K N$ channel in the dynamics together with the scarce influence of the $|qqq\rangle$ component compared to the corresponding one for the charm and bottom resonances. Ref.~\cite{Conde-Correa:2024qzh} contains an interesting analysis of the $\bar{K}N$ system in the framework of CQM, where the two-pole nature of the $\Lambda(1405)$ is only recovered when other meson-baryon channels are taken into account via an optical potential. In Ref.~\cite{Azizi:2023tmw}, the QCD sum rules method was applied and, assuming a molecular pentaquark structure with a $K^--p$ and $\bar{K}^0-n$ admixture, a mass for the $\Lambda(1405)$ of $1406 \pm 128$~MeV was found. 

One of the drawbacks that reduces the restrictive effect of the scattering data at energies slightly above the $K^- p$ threshold is the large uncertainties associated, since the data-taking process went through bubble chambers. In contrast, the precise femtoscopic technique from High-Energy nuclear collisions offers one of the most promising ways to extract information of the hadron-hadron interactions and, in particular, of the $K^- p$ one. The reason lies in the fact that, for these high-multiplicity event reactions, the hadron production yields are well described by statistical models, thus leaving the correlations between the outgoing particles to depend on the final state scattering. This technique is specially welcome to study the interaction of those sectors where there is no chance to perform scattering experiments due to the short lifetime of the particles involved in the initial state. Therefore, the hadron femtoscopy technique provides an unprecedented opportunity to impose constraints on the theoretical models.
In~\cite{ALICE:2019gcn}, the measurement of the $K^- p$ correlation in pp collisions by the ALICE collaboration delivered the first experimental evidence of the opening of the $\bar{K}^0$n channel, showing the sensitivity of this new observable to the underlying coupled-channel dynamics. The latter was investigated for the first time within a dedicated coupled-channel framework by the authors in~\cite{Kamiya:2019uiw}.\\
Recently high-precision correlation data became available also in the $S=-2$ meson-baryon sector with $K^-\Lambda$ pairs~\cite{ALICE:2023wjz}. Actually, in~\cite{Sarti:2023wlg}, a novel method to extract information on $S=-2$ meson-baryon scattering amplitudes was presented employing the $K^- \Lambda$ correlation function (CF) measured by the ALICE collaboration at LHC ~\cite{ALICE:2023wjz}. A similar study was carried out in~\cite{Feijoo:2024bvn} to constrain for the first time the vector meson-baryon scattering amplitude in the $S=0,Q=0$ sector. In line with the former works, yet from the inverse problem perspective, the authors of~\cite{Li:2024tvo} study several $S=-1$ meson-baryon CFs to see how much information can be obtained from them focusing on the $\Sigma^*(1/2^-)$ at the $N \bar K$ threshold. \\

In the present study, we revisit the $K^-p$ CF aiming at demostrating the validity of two benchmark models, developed within the UChPT scheme and constrained to the available scattering data, to reproduce the experimental femtoscopic data. We can say in advance that our conclusions do not support those of \cite{ALICE:2022yyh}, claiming a revision of
the full coupled-channel $K^-p$ interaction models in order to properly describe the measured $K^-p$ CF in relative momentum space obtained in $pp$ collisions at $\sqrt{s}=13$~TeV.
Moreover, we provide novel predictions for the $\pi^-\Lambda$ and $K^+\Xi^-$ CFs, whose future comparison to the corresponding ongoing measurements can provide valuable information about the $S=-1$ meson-baryon interaction not only below the $\bar{K}N$ threshold but also at higher energies. In all cases, we present a detailed analysis of all the physically meaningful elements of the CFs. Finally, we also present two new compatible methods to estimate the production weights in a dedicated Appendix. 

\section{Formalism}
\label{sec:formalism}

\subsection{Femtoscopic correlation function}
\label{subsec:femtoscopy}

In the $S=-1$ sector, different meson-baryon channels with the same quantum numbers couple to each other. In particular, for the charge $Q=0$ case we consider ten channels ($K^-p$, $\bar{K}^0n$, $\pi^0\Lambda$, $\pi^0\Sigma^0$, $\eta\Lambda$, $\eta\Sigma^0$, $\pi^+\Sigma^-$, $\pi^-\Sigma^+$, $K^+\Xi^-$, $K^0\Xi^0$), and for $Q=-1$ we consider six channels ($\pi^-\Lambda$, $\pi^0\Sigma^-$, $\pi^-\Sigma^0$, $K^-n$, $\eta\Sigma^-$, $K^0\Xi^- $). In this multichannel scenario, the two-particle CF of the observed $i$-channel can be expressed through the generalized Koonin-Pratt formula \cite{Koonin1977,Pratt1990}, which has been recently reinforced by the study presented in \cite{Albaladejo:2024lam},
\begin{equation}
    C_i(p) = \sum_{j} w_j \int d^3 r \ S_j(r)|\Psi_{ji}(p,r)|^2\,.
    \label{eq:CF_KP}
\end{equation}
The variables $p$ and $r$ represent the relative momentum and distance between the two particles observed in the pair rest frame, respectively.
The preceding summation covers the possible transitions allowed by the theory at hand from all possible $j$-particle pairs to the final $i$-pair. These $j$-contributions are scaled by the production weights $w_j$, accounting for the amount of primary pairs $j$ produced in the initial pp collision, which can transform into the measured final $i$-pair in a region of $p$ below 300 MeV/c. The calculation of the production weights is performed following the VLC Method described in Appendix~\ref{app:weights}. The corresponding $w_j$'s entering in the different CFs studied in this work are displayed in Table~\ref{tab:production_weights}.

The function $S_j(r)$ stands for the emitting source, and represents the probability distribution of producing the $j$-th pair at a relative distance $r$. For CFs measured in pp collisions at LHC energies, the ALICE Collaboration showed that the emitting source is composed of a Gaussian core, common to all particles, and a non-Gaussian component coming from strongly decaying resonances into the particles forming the pair of interest~\cite{ALICE:2020ibs,ALICE:2023sjd}. In particular, the core source size $r_{\mathrm{core}}$ depends on the average transverse mass $\langle m_{\mathrm{T}}\rangle$ of the pair under study. This core-resonance halo source function is typically parametrized as an effective spherical Gaussian, $S_j(r)=(4\pi R_j^2)^{-3/2}\exp(-r^2/4R_j^2)$, with size $R_j$. The latter can depend on the channel due to the different feed-down of strongly decaying resonances into the particles composing each $j$ pair.\\
For the $K^-p$ CF, we employed the source sizes reported in \cite{ALICE:2022yyh}: $R_{\bar{K}N}=1.08\pm0.18$ fm and $R_{\pi\Sigma,\pi\Lambda}=1.23\pm0.21$ fm\footnote{We report the total uncertainty evaluated as the squared root of the sum in quadrature of the statistical and systematic experimental errors.}. Since in that work the $\eta\Lambda$, $\eta\Sigma^0$, $K^+\Xi^-$ and $K^0\Xi^0$ channels were not included, their corresponding source sizes are fixed to $1.25$ fm. As will be shown later, these channels have a negligible contribution to the $K^-p$ CF and thus the possible variation of their source sizes does not affect our results. For the $K^+\Xi^-$ and $\pi^-\Lambda$ CFs no experimental information is yet available, and the same source size is used for every channel, presenting three cases: $R=1.0$,~$1.25$ and $1.5$~fm.

Finally, the last ingredient in the former expression, $\Psi_{ji}(p,r)$, is the relative wave function for the transition of an intermediate pair channel $j$ to the asymptotically measured one $i$.  Following \cite{Vidaña2023}, this relative wave function can be obtained directly from the scattering amplitude by solving a BS equation. The s-wave component reads:
\begin{eqnarray}
    \varphi_{ji}(p,r) = \delta_{ji}j_0(pr) + \int_0^{q_{\rm max}} \frac{4\pi q^2 dq}{(2\pi)^3} \frac{1}{2\omega_j(q)}\frac{M_j}{E_j(q)} \nonumber \\
    \times \frac{T_{ji}(\sqrt{s},p,q)\ j_0(qr)}{\sqrt{s}-\omega_j(q)-E_j(q)+i\epsilon} \ .
    \label{wf_ampli} 
\end{eqnarray}
Here, $\omega_j(q)=(q^2+m_j^2)^{1/2}$ and $E_j(q)=(q^2+M_j^2)^{1/2}$ are the energies of the meson and the baryon, respectively, in the $j$-th channel, and $j_0(x)$ is the spherical Bessel function of the first kind. The integral is regulated through a cutoff fixed to a natural size $q_{\rm max}=800$ MeV/c. As detailed later, the amplitude $T$ incorporates the strong interaction in s-wave and, in diagonal transitions ($i=j$) for channels involving a pair of charged particles, it also contains the effect of the Coulomb force.

The diagonal wave function component of the emitted pair can be written as:
\begin{equation}
\Psi_{ii}(\boldsymbol{p};\boldsymbol{r})= \Phi_i(\boldsymbol{p};\boldsymbol{r}) - \Phi_{0i}(pr) + \varphi_{ii}(p;r) \ ,  \label{eq:decomp}
\end{equation} 
with 
$\Phi_i(\boldsymbol{p};\boldsymbol{r})=\text{e}^{{\rm i}\boldsymbol{p}\boldsymbol{r}}$ and 
$\Phi_{0i}(pr)=j_0(pr)$ if the particles do not interact electromagnetically. If the observed channel involves a pair of charged particles, the function $\Phi_i$ stands for the complete Coulomb wave function and $\Phi_{0i}$ for its $s$-wave component. The non-diagonal component of wave function $\Psi_{ji}(p,r)$ $(j\ne i)$ is given by the second term on the r.h.s. of Eq.~(\ref{wf_ampli}), where in this case
the amplitude $T_{ji}$ contains only the strong interaction in s-wave.
Taking into account the above considerations, the CF of Eq.~(\ref{eq:CF_KP}) becomes
\begin{eqnarray} 
\label{eq:corr}
&&C_i(p)=\int d^3r \,\, S_i(r) \ |\Phi_i(\boldsymbol{p};\boldsymbol{r}) |^2  \nonumber \\
&+& \int 4\pi r^2 dr \left[ \sum_j S_j(r)  w_j \ | \varphi_{ji}(p;r) |^2 - S_i(r)| \Phi_{0i}(pr) |^2 \right] \ .
\end{eqnarray}

\begin{table*} 
\begin{center}
\begin{tabular}{|c|c|}
\hline
Channel$-j$ & \hspace{0.2cm} $w_j$ ($K^-p$ CF) \hspace{0.2cm} \\ [2mm] \hline
 $K^-p$  & $1$ \\  [1mm]
 $\bar{K}^0n$& $0.97\pm0.20$~\cite{ALICE:2022yyh} \\  [1mm]
$\pi^0\Lambda$& $1.96\pm0.93$~\cite{ALICE:2022yyh} \\ [1mm]
$\pi^0\Sigma^0$& $1.37\pm0.68$~\cite{ALICE:2022yyh} \\ [1mm] 
 $\pi^+\Sigma^-$& $1.42\pm0.71$~\cite{ALICE:2022yyh}\\  [1mm]
$\pi^-\Sigma^+$& $1.41\pm0.70$~\cite{ALICE:2022yyh} \\ [1mm]
 $\eta\Lambda$  & $0$  \\ [1mm]
$\eta\Sigma^0$& $0$ \\ [1mm]
$K^+\Xi^-$& $0$ \\ [1mm]
$K^0\Xi^0$& $0$  \\  [1mm] 
\hline
\end{tabular}
\hspace{0.5cm}
\begin{tabular}{|c|c|c||c|c|}
\hline
Channel$-j$ & \hspace{0.2cm} $w_j$ ($K^-p$ CF) \hspace{0.2cm} & \hspace{0.2cm} $w_j$  ($K^+\Xi^-$ CF) \hspace{0.2cm} &  Channel$-j$ & \hspace{0.2cm} $w_j$ ($\pi^-\Lambda$ CF) \hspace{0.2cm} \\ [2mm] \hline
 $K^-p$  & $1$ & $7.25^{+1.98}_{-1.15}$ &$\pi^-\Lambda$& $1$\\ [1mm]
 $\bar{K}^0n$& $0.90^{+0.09}_{-0.10}$ & $7.20^{+1.75}_{-1.28}$ &$\pi^0\Sigma^-$& $0.45^{+0.03}_{-0.04}$\\  [1mm]
$\pi^0\Lambda$& $1.47^{+0.16}_{-0.14}$ & $8.05^{+2.06}_{-1.26}$ &$\pi^-\Sigma^0$& $0.44^{+0.03}_{-0.03}$\\ [1mm]
$\pi^0\Sigma^0$& $1.15^{+0.13}_{-0.12}$ & $6.42^{+1.69}_{-0.97}$ &$K^-n$& $0.12^{+0.01}_{-0.01}$\\  [1mm]
 $\pi^+\Sigma^-$& $1.00^{+0.13}_{-0.09}$ & $6.17^{+1.46}_{-0.96}$ &$\eta\Sigma^-$& $0$\\  [1mm]
$\pi^-\Sigma^+$& $1.03^{+0.10}_{-0.10}$ & $6.41^{+1.67}_{-1.00}$ &$K^0\Xi^-$& $0$\\ [1mm]
 $\eta\Lambda$  & $0$ & $2.83^{+0.74}_{-0.47}$ &$\_$ & $\_$ \\ [1mm]
$\eta\Sigma^0$& $0$ & $2.03^{+0.60}_{-0.26}$ &$\_$ & $\_$\\
$K^+\Xi^-$& $0$ & $1$ &$\_$ & $\_$\\ [1mm]
$K^0\Xi^0$& $0$ & $1.06^{+0.26}_{-0.18}$ &$\_$ & $\_$ \\ 
[1mm] 
\hline
\end{tabular}
\caption{Values of the production weights for $\pi^-\Lambda$, $K^-p$ and $K^+\Xi^-$ CFs. Left table: old values, taken from Ref.~\cite{ALICE:2022yyh}. Right table: new values following the VLC Method of Appendix~\ref{app:weights}.}
\label{tab:production_weights}
\end{center}
\end{table*} 

\subsection{Strong interaction}
\label{subsec:strong}

In order to model the meson-baryon strong interaction in the $S=-1$ sector, as already mentioned in the Introduction, we employ a UChPT scheme within a coupled-channels formalism.  We obtain the scattering amplitude starting from the effective chiral Lagrangian up to NLO, $\mathcal{L_{\phi B}} = \mathcal{L}^{(1)} + \mathcal{L}^{(2)}$, given by
\begin{eqnarray}
    \mathcal{L}^{(1)} &=& \langle \bar{B} i \gamma^\mu D_\mu B \rangle - M_0 \langle \bar{B}B \rangle + \frac{1}{2} D \langle \bar{B}\gamma^\mu\gamma^5 \{ u_\mu,B \} \rangle \nonumber \\ 
    &+& \frac{1}{2} F \langle \bar{B}\gamma^\mu\gamma^5 [ u_\mu,B ] \rangle \, ,
    \label{eq:lagrangian1}
\end{eqnarray}
\begin{eqnarray}
    \mathcal{L}^{(2)} &=&  b_D \langle \bar{B} \{\chi_+,B\} \rangle + b_F \langle \bar{B}[\chi_+,B] \rangle + b_0 \langle\bar{B}B\rangle \langle \chi_+ \rangle \nonumber \\
    &+& d_1 \langle\bar{B}\{u_\mu,[u^\mu,B]\} \rangle + d_2 \langle \bar{B}[u_\mu,[u^\mu,B]] \rangle  \nonumber \\ 
    &+&  d_3 \langle \bar{B}u_\mu \rangle \langle u^\mu B \rangle  + d_4 \langle \bar{B}B \rangle \langle u^\mu u_\mu \rangle \, .
    \label{eq:lagrangian2}
\end{eqnarray}
Here, $B$ is the octet baryon matrix, while the pseudoscalar mesons matrix $\phi$ is contained in the field $u_\mu = i u^\dagger \partial_\mu U u^\dagger$, where $U = u^2 = \exp(i \sqrt{2} \phi / f)$ and $f$ is the effective meson decay constant. The covariant derivative is defined as $D_\mu B = \partial_\mu B + [\Gamma_\mu, B]$, where $\Gamma_\mu = (u^\dagger \partial_\mu u + u \partial_\mu u^\dagger)/2$, and $\chi_+=-\{\phi,\{\phi,\chi\}\}/4f^2$, where $\chi=\textrm{diag}(m_\pi^2,\ m_\pi^2,\ m_K^2-m_\pi^2)$. This Lagrangian depends on the axial vector constants $D$ and $F$, together with $f$ and the NLO LECs ($b_D,b_F,b_0,d_1,d_2,d_3,d_4$).

\begin{figure*}
    \centering
    \includegraphics[width=0.8 \textwidth]{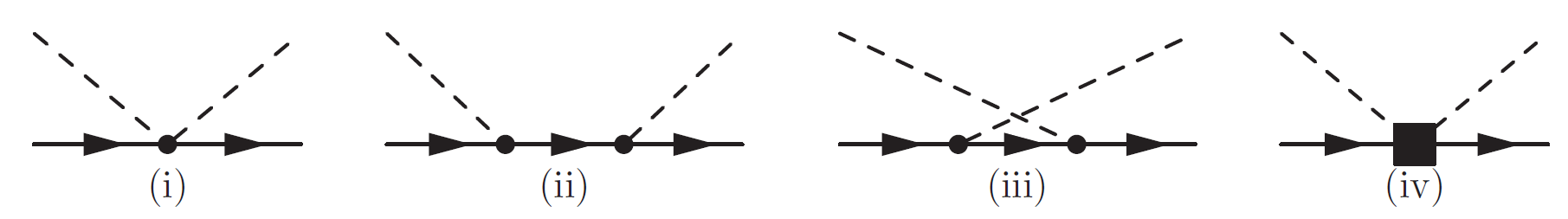}
    \caption{Diagrammatic contributions to the interaction kernel. Contact diagrams (i) and (iv) represent the WT and NLO terms, respectively. Diagrams (ii) and (iii) represent the direct Born (s-channel) and the crossed Born (u-channel) terms, respectively.}
    \label{fig:interactionkernel}
\end{figure*}

The interaction kernel is obtained from Eqs.~(\ref{eq:lagrangian1}) and (\ref{eq:lagrangian2}) projecting the resulting potential onto its $L=0$ partial wave and averaging over polarization states,
\begin{equation}
    V_{ij}(\sqrt{s}) = \frac{1}{8\pi} \sum_{\sigma,\sigma'} \int d\Omega \ \hat{V}_{ij}(\sqrt{s},\Omega,\sigma,\sigma')
\end{equation}
where $V_{ij}$ represent the different interaction kernel matrix elements associated to the transitions from the incoming i-th to the outgoing j-th channels. The interaction kernel can be separated into four contributions: the Weinberg-Tomozawa (WT) calculated from the covariant derivative, the direct Born and crossed Born terms, whose vertices are obtained from the last two terms of $\mathcal{L}^{(1)}$, and the NLO terms at tree level, derived from $\mathcal{L}^{(2)}$. These contributions are depicted diagrammatically in Fig.~\ref{fig:interactionkernel}. For more details on the explicit form of these contributions, see \cite{Ramos:2016odk,Feijoo:2021zau}.

Unitarity is implemented into the scheme via the BS equations, whose solution leads to scattering amplitudes. Within the on-shell approximation~\cite{OR,HJ_rev} the BS equations can be written in the matrix form
\begin{equation}
    T = (1-VG)^{-1}V
\end{equation}
where $G$ is a diagonal matrix whose elements are the meson-baryon loop functions, which diverge logarithmically and, therefore, have to be regularized. Following the dimensional regularization (DR) method, its diagonal elements can be expressed as  
\begin{eqnarray}
    G_l(\sqrt{s}) &=& \frac{2M_l}{16\pi^2} \bigg[ a_l(\mu) + \ln\frac{M_l^2}{\mu^2} + \frac{m_l^2-M_l^2 + s}{2s}\ln\frac{m_l^2}{M_l^2} \nonumber \\
    &+& \frac{q_l}{\sqrt{s}} \ln\frac{(s+2q_l\sqrt{s})^2 - (M_l^2-m_l^2)^2}{(-s+2q_l\sqrt{s})^2 - (M_l^2-m_l^2)^2} \bigg]
\end{eqnarray}
where $M_l$ and $m_l$ are, respectively, the baryon and meson masses of the l-th channel and $q_l$ the center of mass (CM) momentum of the l-th channel pair at a $\sqrt{s}$ energy. The $a_l$ are the so-called subtraction constants (SCs), which replace the divergence for a given regularization scale $\mu$. In principle, there are as many SCs as channels taken into account in the considered sector, but the number of SCs can be reduced assuming isospin-symmetry arguments for the elements of each channel multiplet.

Throughout this work we employ two models. On the one hand, the Oset-Ramos model \cite{2pole}, which is based on an interaction kernel consisting of the WT contribution and is regularized via DR thereby depending on 7 free parameters, namely the decay constant $f$ and 6 SCs (one for each meson-baryon channel, taking isospin symmetry into account). On the other hand, the BCN model, which corresponds to the WT+Born+NLO fit in \cite{Feijoo:2018den} and employs the full interaction kernel from Eqs.~(\ref{eq:lagrangian1}) and (\ref{eq:lagrangian2}), and thus depends on 16 free parameters (10 LECs and 6 SCs). Both models are limited to s-wave contributions.

\subsection{Coulomb interaction}
\label{subsec:coulomb}

For the $Q=0$ case, besides the strong interaction, one must take into account the Coulomb force in the channels involving a pair of charged particles. Following the procedure described in Ref.~\cite{Holzenkamp:1989tq}, successfully applied in the femtoscopy study of Ref.~\cite{TorresRincon2023}, the Coulomb amplitude
is obtained by Fourier transforming the Coulomb potential to momentum space:
\begin{eqnarray}
V^{\rm c}(\varepsilon,| \boldsymbol{p}^\prime- \boldsymbol{p} |) &=&  \int^{R_c}_{0} d^3r \ {\rm e}^{ i (\boldsymbol{p}^\prime-\boldsymbol{p}) \cdot \boldsymbol{r} } \ \frac{\varepsilon\alpha}{r} \nonumber \\
 &=& \frac{4\pi \varepsilon \alpha}{| \boldsymbol{p}^\prime- \boldsymbol{p}  |^2}\left[ 1-\cos( |\boldsymbol{p}^\prime- \boldsymbol{p}  | R_c)\right] \ ,
 \label{eq:coul}
\end{eqnarray} 
where $\varepsilon=+1 (-1)$ indicates the repulsive (attractive) character of the Coulomb interaction between the pair of charged particles, $\alpha$ is the fine structure constant and $R_c=60$ fm is a regulator introduced to truncate the Fourier transform of the Coulomb potential and thus avoid the divergence at $\bm{p'}=\bm{p}$. In order to combine it with the s-wave strong interaction, we need to extract the s-wave component of the Coulomb amplitude, which reads:
\begin{eqnarray}
    V^c(\varepsilon,p,p') &=& \frac{2\pi\epsilon\alpha}{pp'} \bigg\{ Ci[|p'-p|R_c] - Ci[(p'+p)R_c] \nonumber \\
    &+& \ln\frac{(p'+p)R_c}{|p'-p|R_c} \bigg\} \ ,
\end{eqnarray}
where $Ci[x]=\int_x^\infty dt \cos(t)/t$ is the cosine integral function. This combination also requires modifying the Coulomb amplitude with some factors, namely:
\begin{eqnarray}
    \tilde{V}^c(\sqrt{s},p,p') &=& \sqrt{\frac{E(p)}{M}}\sqrt{2\omega(p)}\sqrt{\xi(p,s)}\times V^c(\varepsilon,p,p') \nonumber \\
    &\times& \sqrt{\xi(p',s)}\sqrt{2\omega(p')}\sqrt{\frac{E(p')}{M}}
\end{eqnarray}
where $E$ and $M$ are the energy and mass of the baryon, $\omega$ the energy of the meson and 
\begin{equation}
    \xi(p,s) = \frac{\sqrt{s} - E(p) - \omega(p)}{\tilde{p}^2/2\mu - p^2/2\mu} \ ,
\end{equation}
with $\mu$ being the pair's reduced mass, $\sqrt{s}$ the CM energy and $\tilde{p}$ the corresponding on-shell momentum. These factors are needed because the static Coulomb potential $\varepsilon\alpha/r$ produces the well known Coulomb wave-functions when inserted into a non-relativistic Lippmann-Schwinger type propagator, whereas we employ a relativistic propagator and relativistic meson and baryon normalization factors in the BS equation.

A common procedure would consist in unitarizing simultaneously the combined strong and Coulomb amplitudes, within a cut-off regularization scheme, as done in \cite{TorresRincon2023}. However, the strong interaction models explored in the present work have been fitted to the scattering observables employing the DR unitarization scheme. Applying a cut-off unitarization scheme would modify the predictions of the scattering observables, hence requiring a refitting of the models, which is not the purpose of the present work. We recall that we aim at exploring the ability of the meson-baryon interaction models (constrained to the scattering data in the strangeness $S=-1$ sector) in reproducing the $K^-p$ correlation function, as well as showing predictions for the $\pi^-\Lambda$ and $K^+\Xi^-$ correlation functions in the same sector. Therefore, to avoid modifying the already tuned strong amplitudes, we opt for obtaining the total scattering amplitude as
\begin{eqnarray}
    T_{ij}(\sqrt{s},p,p') = T_{ij}^S(\sqrt{s}) + \delta_{ij} T^c_{i}(\sqrt{s},p,p')
    \label{eq:vs+vc}
\end{eqnarray}
where $T^S$ is the strong scattering amplitude obtained in section \ref{subsec:strong} and $T^c(\sqrt{s},p,p')=\tilde{V}^c(\sqrt{s},p,p')$ is the Coulomb one up to first order in the BS equation\footnote{We have numerically checked that solving the BS equation with the combined strong and Coulomb interactions (employing cut-off values in the range $600-1200$ MeV) gives a CF that does not differ much from the sum of the separatedly unitarized amplitudes shown in Eq.~\ref{eq:vs+vc}, the difference lying well within the error bands explored in this work. We have also checked that the first order Coulomb amplitude is practically indistinguishable from
the fully unitarized one.}. 


\section{Results}
\label{sec:results}

In this section we present the results obtained for each CF separately. Apart from providing a prediction for the different CFs employing the Oset-Ramos and BCN models, we delve deeper into the relevance of the contributing terms in the interaction kernel. Furthermore, we show the role of each transition wave function $\Psi_{ji} (p,r)$ for the corresponding CF. We start computing $C_{i}(p)$ taking into account only the elastic $\Psi_{ii}$ and, following Eq.~(\ref{eq:CF_KP}), we progressively add the other channel contributions. Despite the Oset-Ramos model provides an overall good description of the available $\bar{K}N$ data, a special attention is paid to the BCN model given the great accuracy shown in describing such data thanks to the incorporation of higher order corrections in the kernel.

\subsection{$K^-p$ correlation}

For a fair comparison to the study of \cite{ALICE:2022yyh}, we start discussing what one would get using the same inputs employed there, apart from the wave functions, which are here replaced by those obtained from the BCN model. Afterwards, we implement more accurate weights, calculated employing the VLC Method described in Appendix~\ref{app:weights}, which provides more reliable results. Compared to the procedure described in
Ref.~\cite{ALICE:2022yyh}, the 
methods developed in this work for obtaining the production weights treat the coupled channel dynamics more consistently and, in the case of the VLC Method,  a pseudorapidity distribution more in accordance with the assumed thermal emission is employed (see Appendix~\ref{app:weights} for details).

\begin{figure}
    \centering
    \includegraphics[width=\columnwidth]{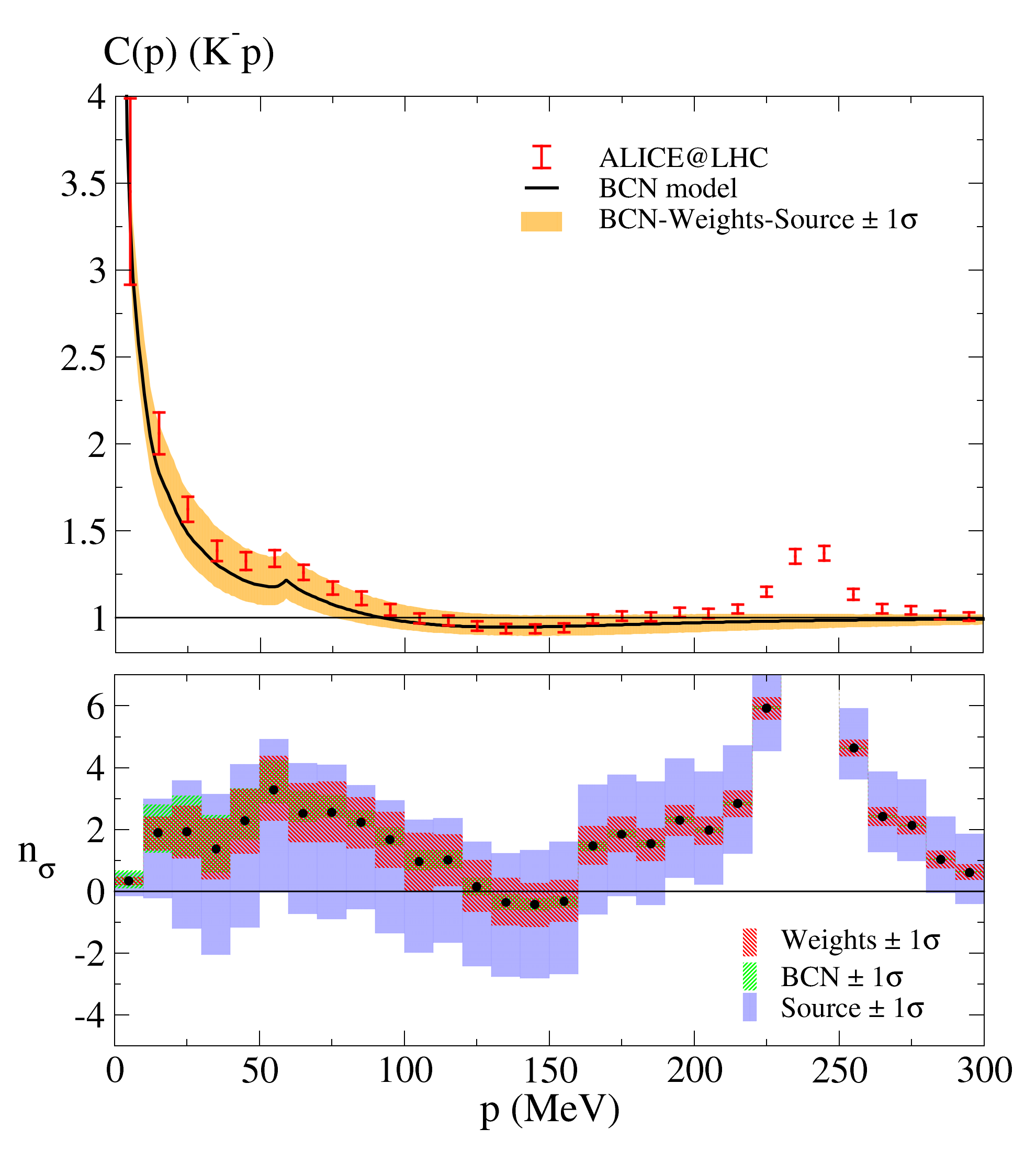}
    \caption{Upper plot: $K^-p$ CF predicted by the BCN model  (black line) when using the production weights used in \cite{ALICE:2022yyh} (left panel in Table~\ref{tab:production_weights}). The experimental data points are taken from~\cite{ALICE:2022yyh} ($p-p$ collision data set at $\sqrt{s}=13$ TeV in Fig.4). The error band is derived from all sources of uncertainty combined. Lower plot: relative deviation between the model prediction and the experimental data, $n_\sigma=(C_{exp}-C_{model})/\sigma_{exp}$, showing separately the different sources of uncertainty.}
    \label{fig:OLD_CF}
\end{figure}

Fig.~\ref{fig:OLD_CF} contains the $K^-p$ CF obtained from the BCN model combined with the Coulomb interaction and taking the production weights from \cite{ALICE:2022yyh} (see first panel of Table~\ref{tab:production_weights}). The model shows a nice  agreement with the CF data in the $p$ range below $\approx 220$ MeV/c. The structure around $p\approx 250$ MeV/c corresponding to the $\Lambda(1520)$, with $J^P=3/2^-$, cannot be reproduced by the bare theoretical models since they are limited to s-wave. We can also observe that in the region around the $\bar{K}^0n$ cusp, the BCN model has a slightly lower strength, bringing it further away from the data points. However, within the total associated errors, including the data points, LECs, source sizes and weight uncertainties\footnote{To estimate the error bands we randomly sample the values of the parameters needed to calculate the CF (LECs, SCs, $w_j$'s and $R_j$'s) assuming they are gaussianly distributed. We iterate the process $~10^3-10^4$ times and, for each parameter set, we calculate the corresponding CF. Afterwards, from the normal distribution of CF values around each momentum, we extract the associated standard deviation.}, we can appreciate that the BCN model is able to reproduce the data within a maximum of 1$\sigma$ deviation (orange band). In order to understand in more detail the impact of the different sources of uncertainty, we present in the lower panel of Fig.~\ref{fig:OLD_CF} the deviation between model and data expressed in terms of the number of standard deviations $n_\sigma$.
Excluding the region of the $\Lambda(1520)$ for the reasons stated above, the BCN model shows the largest deviation with data in the momentum range from $45$ to $90$~MeV/c. The uncertainty related to the LECs (green boxes) shows a significant reduction in the first entry given the restrictive constraints provided by the observables at $\bar{K}N$ threshold (the precise kaonic hydrogen measurement and the branching ratios) used in the fitting procedure to obtain the BCN Model. This is followed by a region, ranging from $10$ to $80$~MeV/c, where such uncertainties take larger values, being a clear reflection of the sizable uncertainties related to the cross-section data in this region used for the derivation of the model. From $90$~MeV/c on, the relevance of the LECs is progressively reduced in the determination of the global uncertainty, leaving the emitting source size and weights uncertainties as the dominant contributions. In particular, the error associated to the source sizes $R_j$ is the main contribution to the uncertainty of the CFs, as observed in previous femtoscopic analyses~\cite{ALICE:2021njx,ALICE:2020ibs,ALICE:2023sjd}.\\
In Ref.~\cite{ALICE:2022yyh}, for the analogous case in $p-p$ collisions, the authors compare a theoretical $K^-p$ CF, calculated from the chirally motivated potentials of~\cite{Miyahara:2018onh}, with the same experimental data we report in Fig.~\ref{fig:OLD_CF}. By assuming the same production weights displayed in the first panel of Tab.~\ref{tab:production_weights} and the same sources mentioned in Sec.~\ref{sec:formalism}, the modeled CFs largely underestimate the data in a much wider momentum range $[20,90]$~MeV/c (blue band in Fig. 4 of Ref.~\cite{ALICE:2022yyh}). A better description of the data is achieved only if the correlation term $C_{\bar{K}^0n}=w_{\bar{K}^0n}\int S_{\bar{K}^0n} |\Psi_{\bar{K}^0n\rightarrow K^-p}|^2 d^3r$ entering in Eq.~(\ref{eq:CF_KP}) is enhanced by roughly a factor 2 (red band in Fig. 4 of Ref.~\cite{ALICE:2022yyh}).  This has been interpreted by the authors as the need for a possible re-tuning of the $\bar{K}^0 n$  channel-coupling strength in the Kyoto-Munich chiral model. It has to be stressed that the approach adopted in Ref.~\cite{ALICE:2022yyh}, involving just a rescaling of the inelastic correlation terms, was an effective way to gauge the necessity to revise the coupled-channel $K^-p$ amplitude.
Our findings, as seen already in Fig.~\ref{fig:OLD_CF}, show that within the current uncertainty related to the measurements and to the LECs, the BCN model is able to reproduce the cusp region of the CF, showing the compatibility between scattering and CF data in an indirect way.
\begin{figure}
    \centering
    \includegraphics[width=\columnwidth]{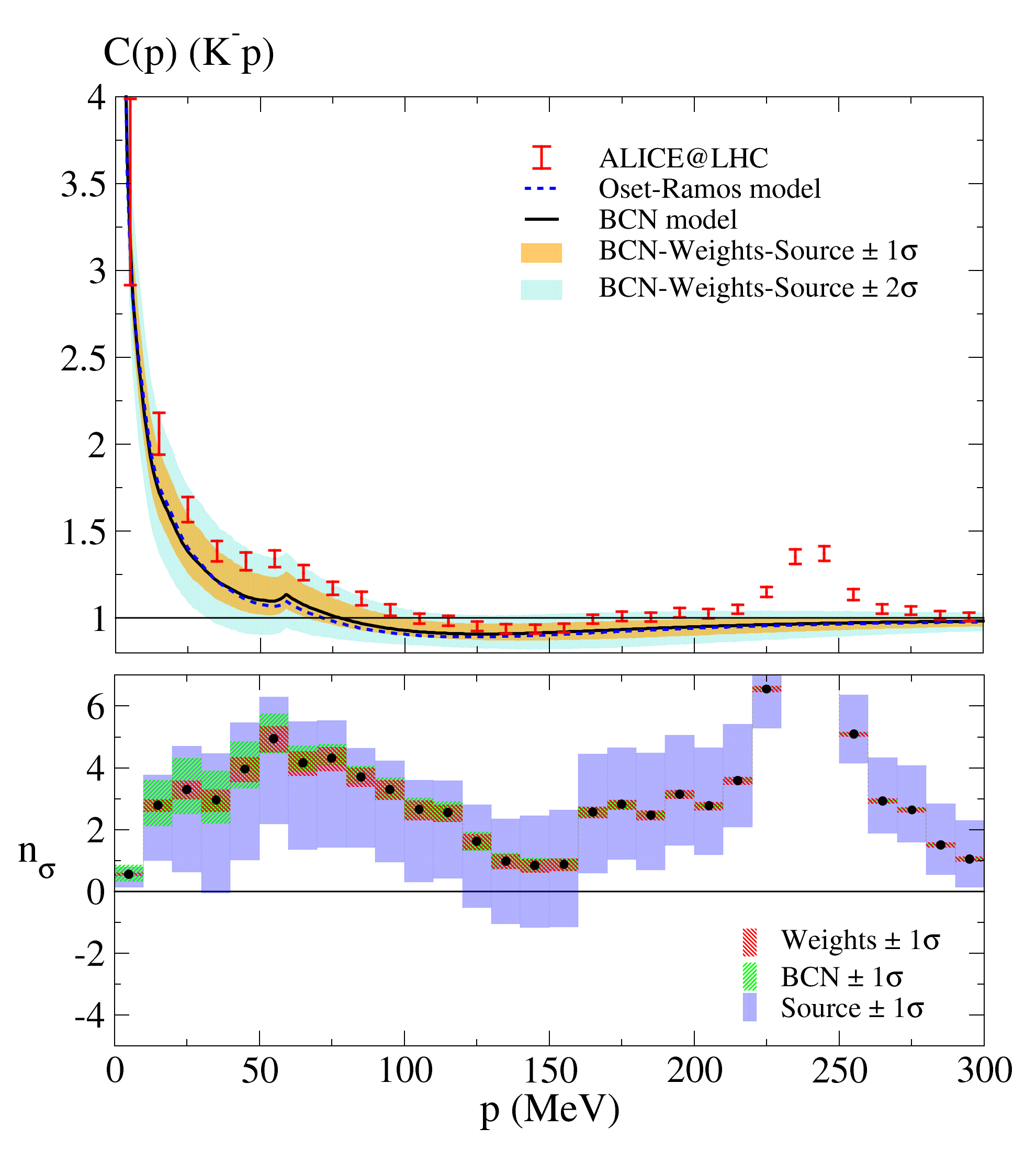}
    \caption{Upper plot: $K^-p$ CFs for the BCN and Oset-Ramos models (black and blue lines respectively), as well as the error bands associated to BCN model (see details in the text). The experimental data points are taken from~\cite{ALICE:2022yyh} ($p-p$ collision data set at $\sqrt{s}=13$ TeV in Fig.4). Lower plot: relative deviation between the model prediction and the experimental data, $n_\sigma=(C_{exp}-C_{model})/\sigma_{exp}$, showing separately the different sources of uncertainty.}
    \label{fig:Full_CF}
\end{figure}

In the same vein, yet using more reliable $w_j$'s obtained with the VLC Method (second column of right hand panel in Table~\ref{tab:production_weights}), we show the $K^-p$ CFs obtained employing Oset-Ramos and BCN models in Fig.~\ref{fig:Full_CF}. As expected from the reduction in the $w_j$ values, both models, which present a very similar behavior, show a slightly worse reproduction of the experimental data. However, such discrepancies lie well within a 2$\sigma$ uncertainty band as can be appreciated from the light blue band in Fig.~\ref{fig:Full_CF}, which is associated to the BCN model and estimated in the same way explained above. This is also reflected in the bottom panel of this figure, which qualitatively follows the analogous one of Fig.~\ref{fig:OLD_CF}. The outputs presented in the current figure reinforce even more the conclusion stated in the previous paragraph regarding the lack of tension between the theoretical models constrained by scattering experiments and the femtoscopic data. \\

\begin{figure}
    \centering
    \includegraphics[width=\columnwidth]{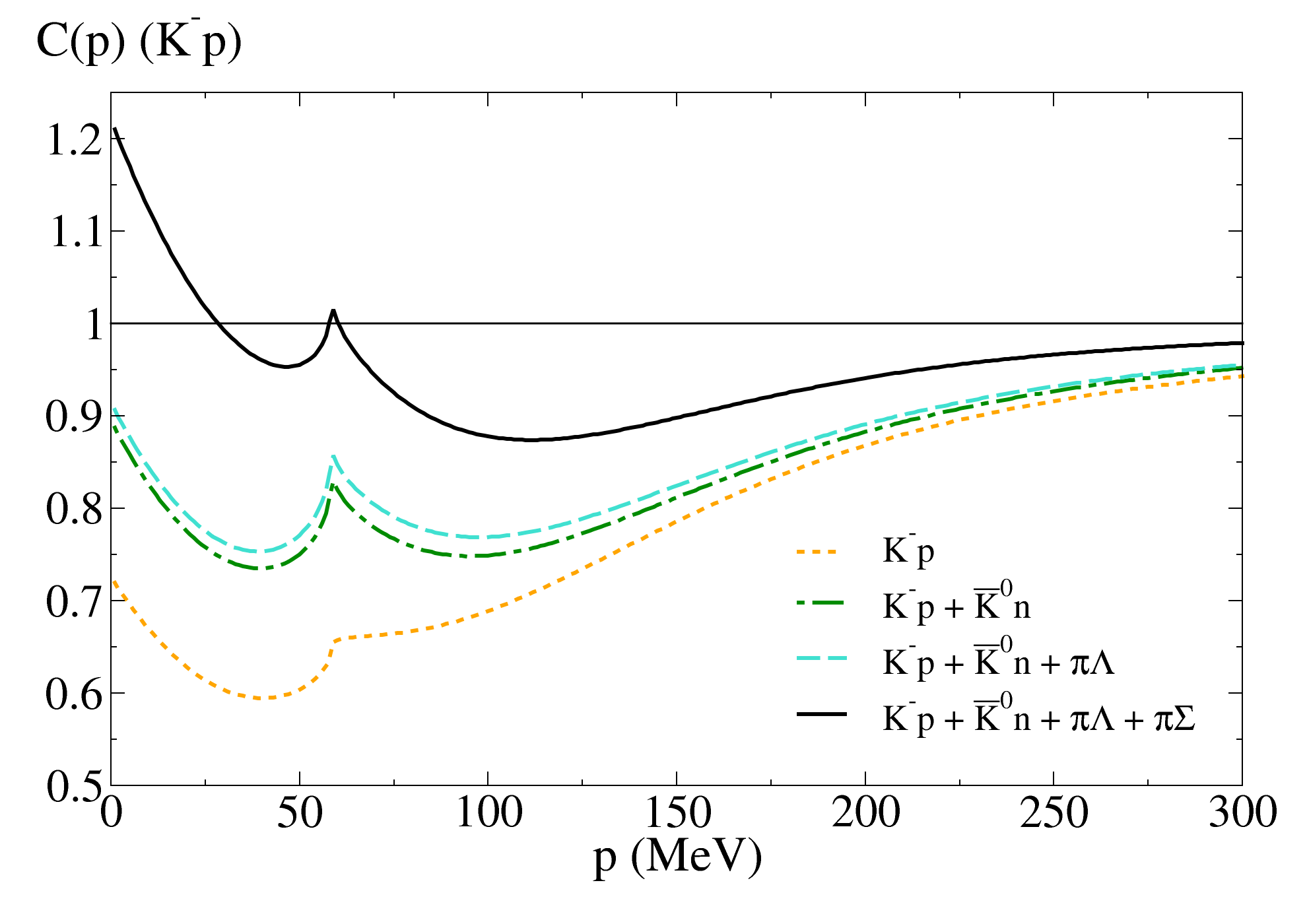}
    \caption{Contribution of the different transitions to the $K^-p$ CF.}
    \label{fig:cc}
\end{figure}

Continuing the CF analysis with the production weights of the VLC Method, we show in Fig.~\ref{fig:cc}  the relevance of coupled-channel in the $K^-p$ CF, omitting Coulomb for this discussion. One immediately realizes that the elastic transition, displayed by an ochre line, would not able to reproduce by itself the experimental data (see Fig.~\ref{fig:Full_CF}) even after implementing the Coulomb effects. From Fig.~\ref{fig:cc}, it can also be noted that the $\bar{K}^0n,\pi\Lambda,\pi\Sigma$ channels contribute to the total $C_{K^- p}(p)$ to a greater or lesser extent depending on the importance of their transition amplitude to the measured $K^-p$ channel, as well as on the value of their production weight. The vanishing production weights of the heavier channels (see the second column of the right hand panel in Table~\ref{tab:production_weights}), make the $\eta\Lambda,\eta\Sigma,K\Xi$ inelastic transitions to be irrelevant for this observable. One of the achievements of the measured $K^-p$ CF~\cite{ALICE:2019gcn} is the observation of a structure around a relative momentum of $p=58$~MeV/c that constitutes the first experimental evidence for the opening of the $\bar{K}^0n$ channel. This signature cusp structure is clearly seen in the different contributions of Fig.~\ref{fig:cc} as a clear consequence of the coupled-channel effects arising from the unitarization.
\begin{figure}
    \centering
.    \includegraphics[width=\columnwidth]{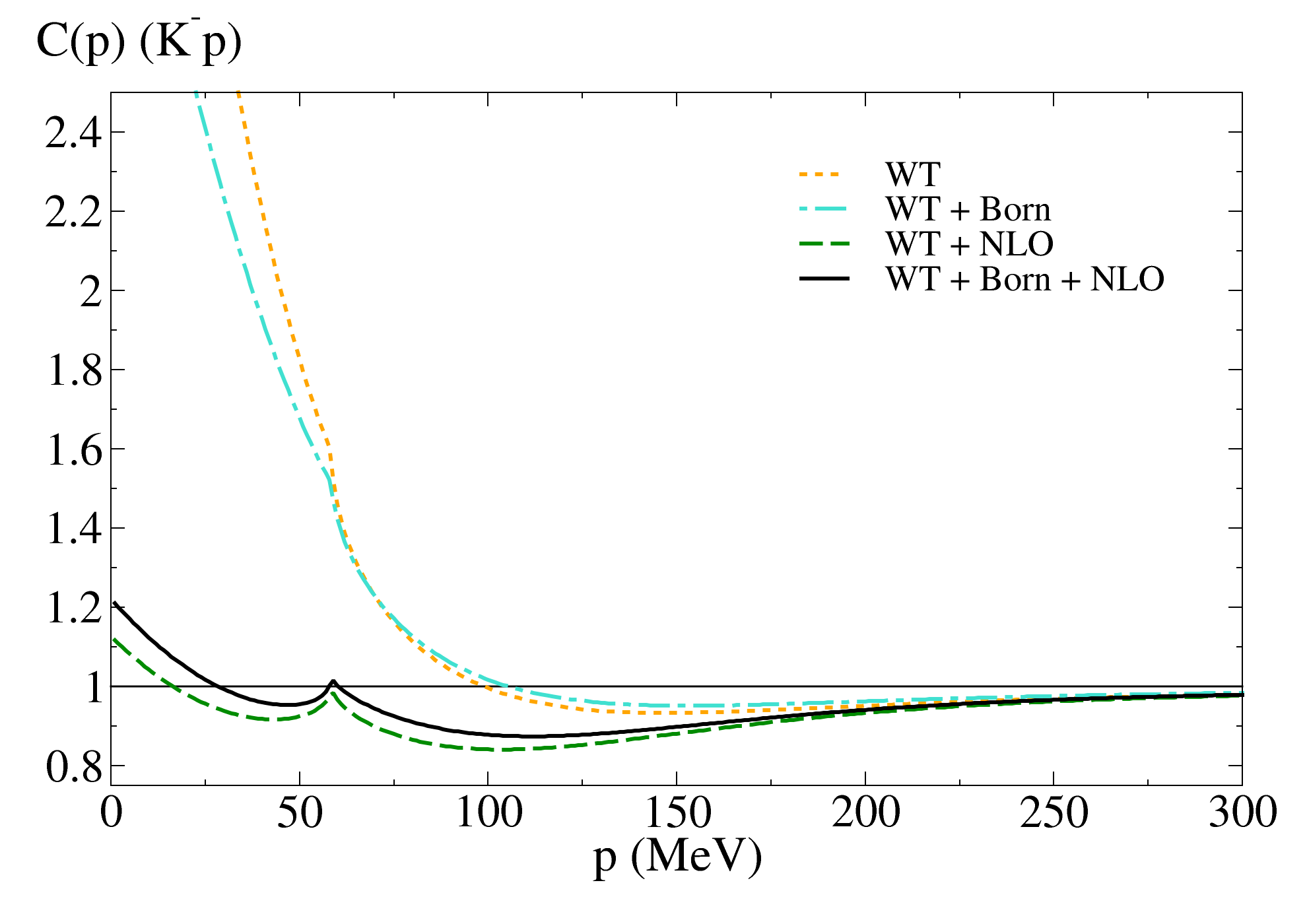}
    \caption{Contribution of the different interaction-kernel terms into the $K^-p$ CF.}
    \label{fig:kernel_terms}
\end{figure}

In Fig.~\ref{fig:kernel_terms}, the role of the different pieces contributing to the interaction kernel of the BCN model is addressed. The green line, which represents what one gets when dealing with an interaction kernel obtained out of the WT+NLO terms, shows the dominance of both former terms in $C_{K^- p}(p)$. The comparison between the black line, accounting for the full kernel, and the green one reveals the mere fine tuning role of the Born terms for this CF. The most eye-catching feature of the plot is the large difference once the NLO contributions are incorporated. This effect is explained by the fact that the parameters of the BCN model were obtained by taking into account all the contributions (WT+Born+NLO) in the fitting procedure. Here, by switching off one of such terms, one is just using an incomplete model. This means that the threshold observables cannot be properly reproduced and the $\Lambda(1405)$ is shifted according to the attractive or repulsive character of the remaining corrections. Actually, the global effect of not accounting for the NLO corrections is a $30$~MeV shift of the $\Lambda(1405)$ pole towards higher energies (settling it at around $1435$~MeV). For the latter case, the strength of all the amplitudes ($T_{j,K^-p}$) entering in the Koonin-Pratt formula (Eq. (\ref{wf_ampli})) get enhanced at threshold, an effect which is reflected in the WT+Born curve (cyan line). Although it seems a counterintuitive interference pattern, since a simple addition of corrections to the dominant contact term is expected, the threshold enhancement found here shows the intricate mechanism behind the UChPT formalism in the presence of a dynamically generated resonance.

\subsection{$\pi^-\Lambda$ correlation}
\begin{figure}
    \centering
    \includegraphics[width=\columnwidth]{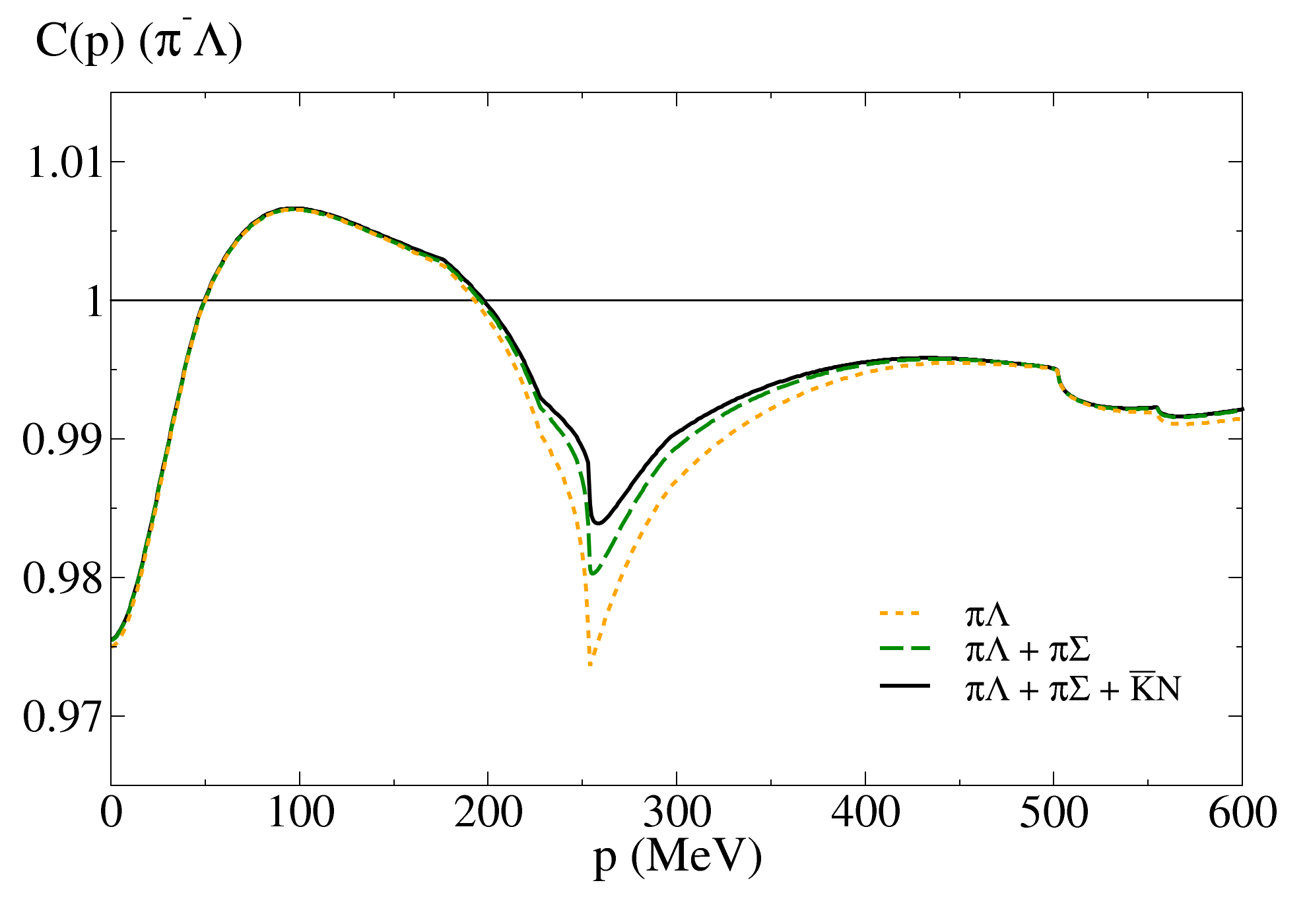}
    \caption{Contribution of the different transitions to the $\pi^-\Lambda$ CF for a source size $R=1.25$ fm.}
    \label{fig:PL_chan}
\end{figure}
We now analyze the $\pi^-\Lambda$ CF following the same scheme as in the previous case. Before moving on in describing the contribution of the different channel transitions, shown in Fig.~\ref{fig:PL_chan}, it is worth reminding that this CF, once measured, can provide novel information about $\bar{K}N$ subthreshold amplitudes. An important fact to be considered is the null elastic transition for the dominant contact term. Actually, the only surviving WT contributions are the $K^-n \to \pi^-\Lambda$ and $K^0 \Xi^- \to \pi^-\Lambda$ transitions (as can be inferred from Table~I in \cite{OR}). Apart from the region around $K^-n$ opening, the dominance of the elastic $\pi^-\Lambda$ transition is shown as main feature when inspecting Fig.~\ref{fig:PL_chan}. This opening appears as an inverted cusp due to a destructive interference pattern, whose effect is reversed by the addition of the inelastic transitions. Apart from the above-mentioned effect at the $K^-n$ opening threshold, we can appreciate the cusps associated to the opening of the  $\eta \Sigma^-$ and  $K^0 \Xi^-$ channels. On the other hand, there is no trace of the $\pi \Sigma$ openings because the $\pi^-\Lambda  \to \pi \Sigma$ transitions can only proceed via the Born terms which provide a negligible contribution around the $\pi \Sigma$ threshold \footnote{The null contribution of the WT and the NLO terms in the $\pi^-\Lambda  \to \pi \Sigma$ transitions can be explained by the zero value of the corresponding couplings, as shown in Tables VII and VIII of Ref.~\cite{Feijoo:2015yja}}.     

\begin{figure}
    \centering
    \includegraphics[width=\columnwidth]{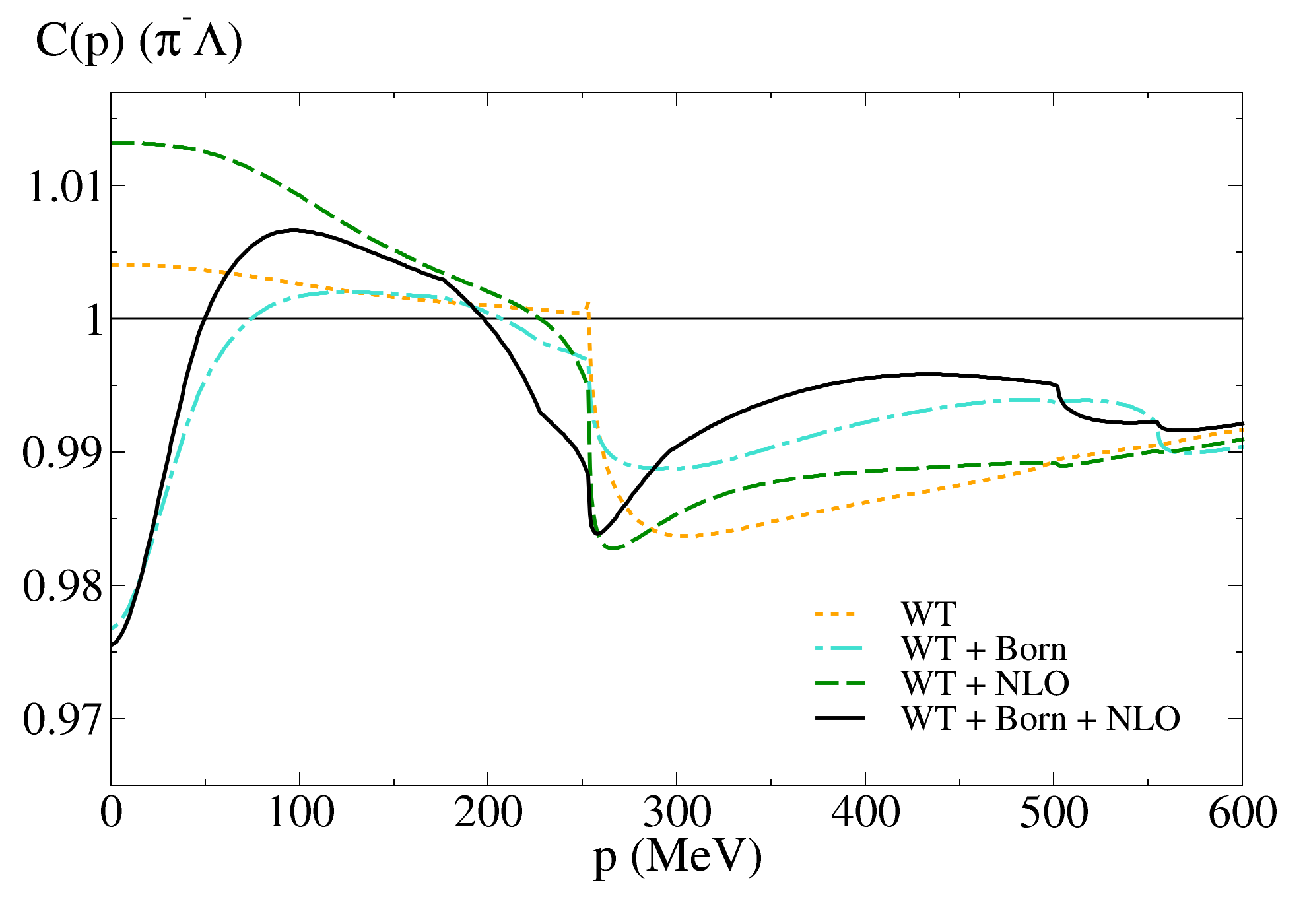}
    \caption{Contribution of the different interaction-kernel terms to the $\pi^-\Lambda$ CF for a source size $R=1.25$ fm.}
    \label{fig:PL_potential}
\end{figure}

 The contribution to the $\pi^-\Lambda$ CF from the different terms of the interaction kernel is displayed in Fig.~\ref{fig:PL_potential}. The limited relevance of the WT terms confers a dominant character to the Born and NLO terms as well as the coupled-channel effect. This is the reason for the small deviation from one shown by the $\pi^-\Lambda$ CFs obtained for the different combinations of the interaction kernel terms. Upon inspecting the low momentum region, one observes a sizable contribution from the Born terms evidenced by the reduction of the black line (obtained with the full kernel) with respect to the green line (obtained when building the interaction from the WT+NLO terms). 

\begin{figure}
    \centering
    \includegraphics[width=\columnwidth]{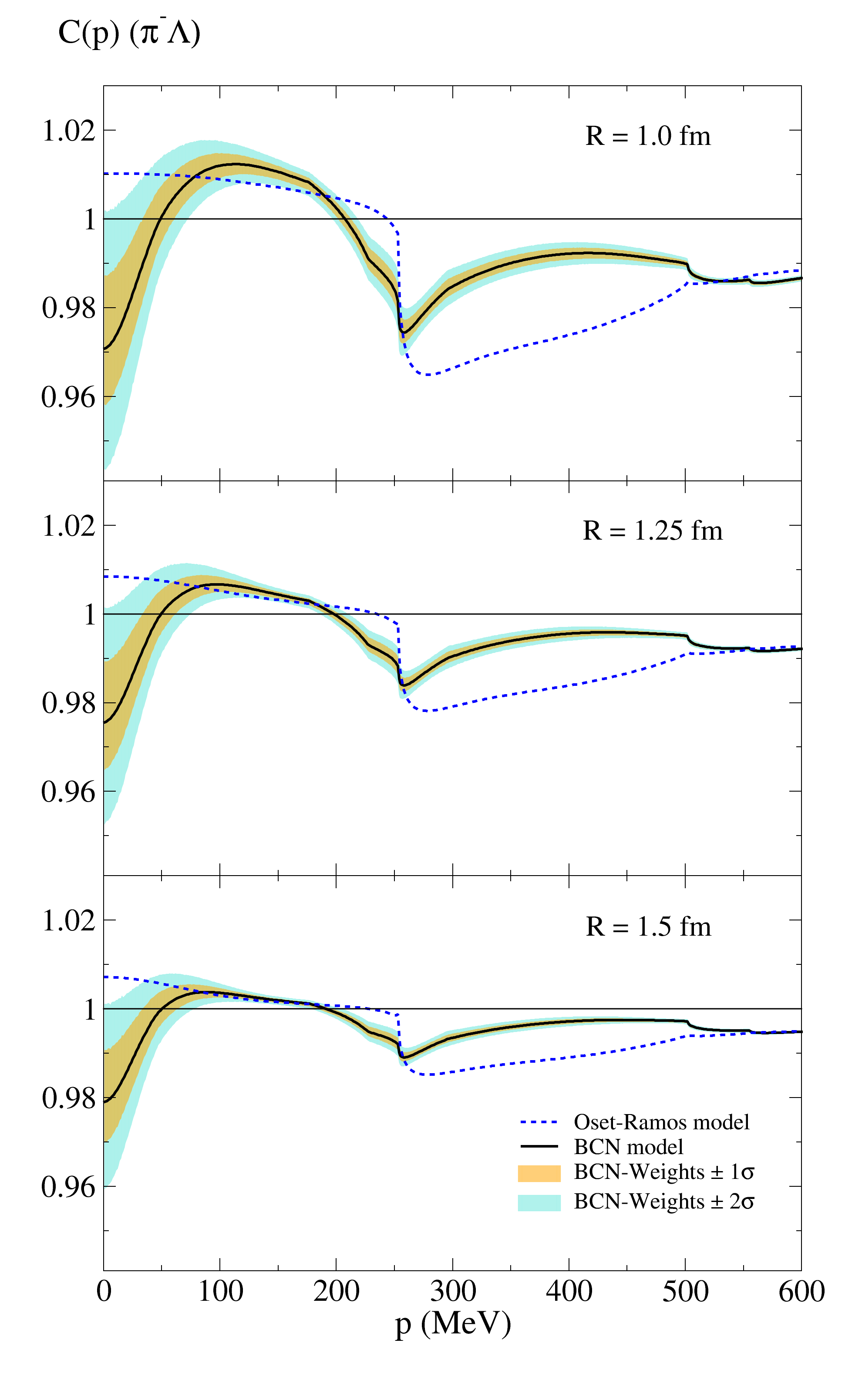}
    \caption{$\pi^-\Lambda$ CFs for the BCN and Oset-Ramos models (black and blue lines respectively), as well as the error bands associated to BCN model (see details in the text), for three values of the source size $R$.}
    \label{fig:PL_BCN_OR}
\end{figure}

Finally, in Fig.~\ref{fig:PL_BCN_OR}, we display the comparison of the obtained $\pi^-\Lambda$ CFs employing the BCN and Oset-Ramos models for different source size radii ($R=1,\, 1.25,\, 1.5$~fm). The null contributions of many channel transitions to the WT term, which is the only one considered in the Oset-Ramos model, makes the corresponding CF to be quite featureless. In contrast, the BCN model has a richer pattern of contributions intertwined by the coupled-channel formalism, as already said above, hence producing a substantial drop in the $\pi^-\Lambda$ CF at low relative momenta.  We can also observe that, as the source size increases, both models acquire a smoother behavior and the discrepancies between them become progressively smaller as expected. We also include the quite sizable error bands corresponding to the BCN model, which makes the discrepancies between both models  at low momenta to be less relevant. 
 
\subsection{$K^+\Xi^-$ correlation}

\begin{figure}
    \centering
    \includegraphics[width=\columnwidth]{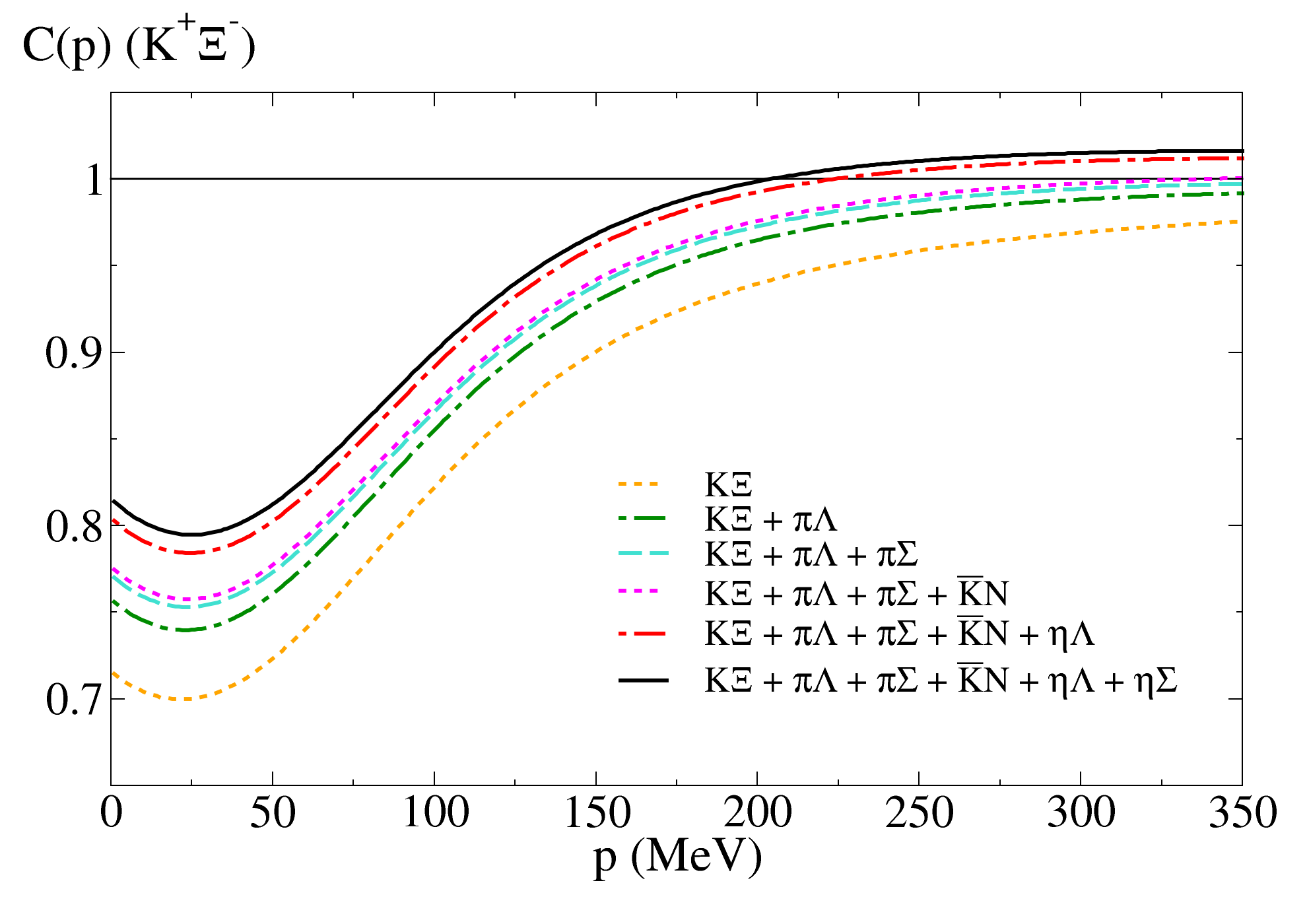}
    \caption{Contribution of the different transitions to the $K^+\Xi^-$ CF for a source size $R=1.25$ fm.}
    \label{fig:KX channels}
\end{figure}
\begin{figure}
    \centering
    \includegraphics[width=\columnwidth]{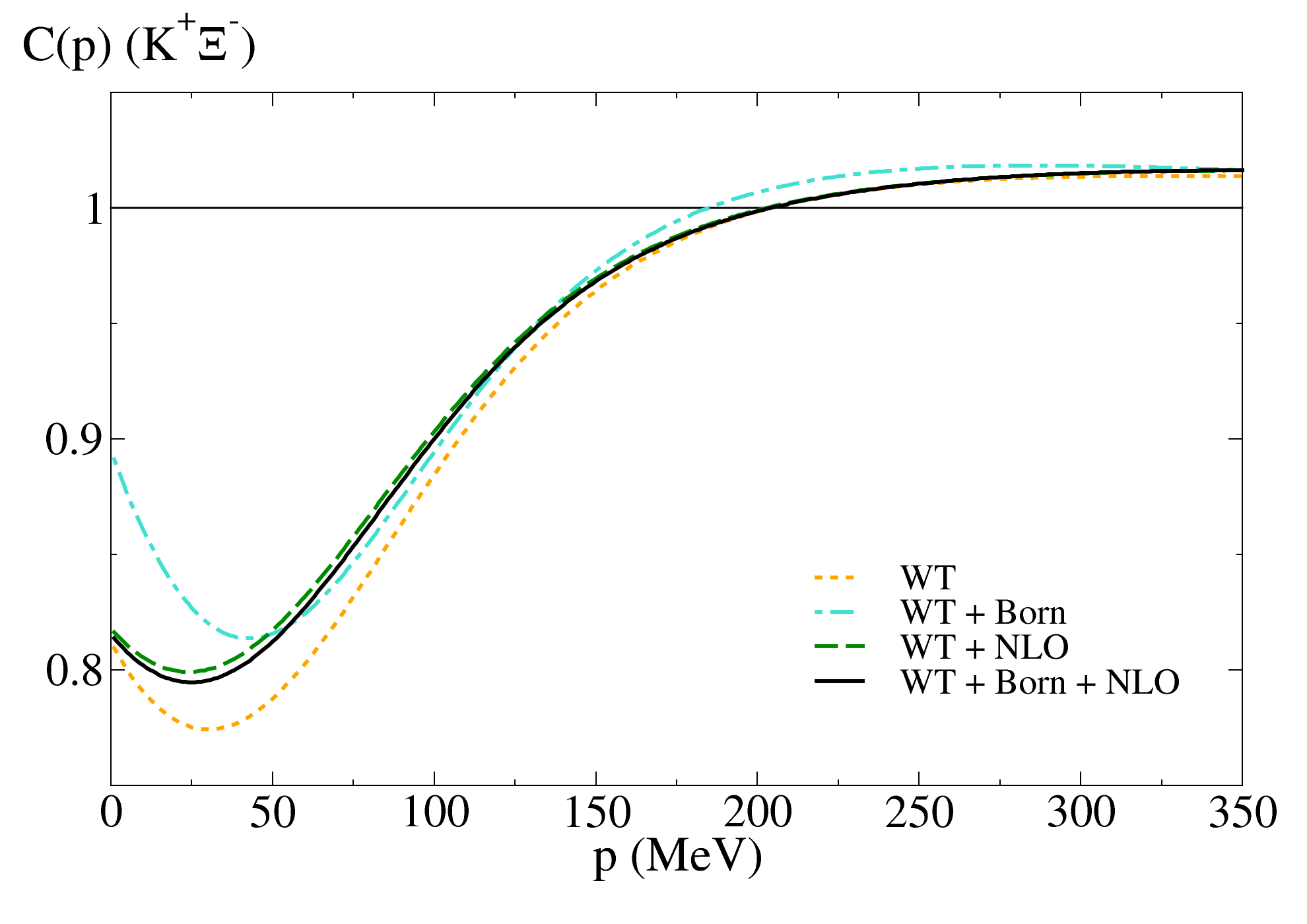}
    \caption{Contribution of the different interaction-kernel terms to the $K^+\Xi^-$ CF for a source size $R=1.25$ fm.}
    \label{fig:KX kernel}
\end{figure}
\begin{figure}
    \centering
    \includegraphics[width=\columnwidth]{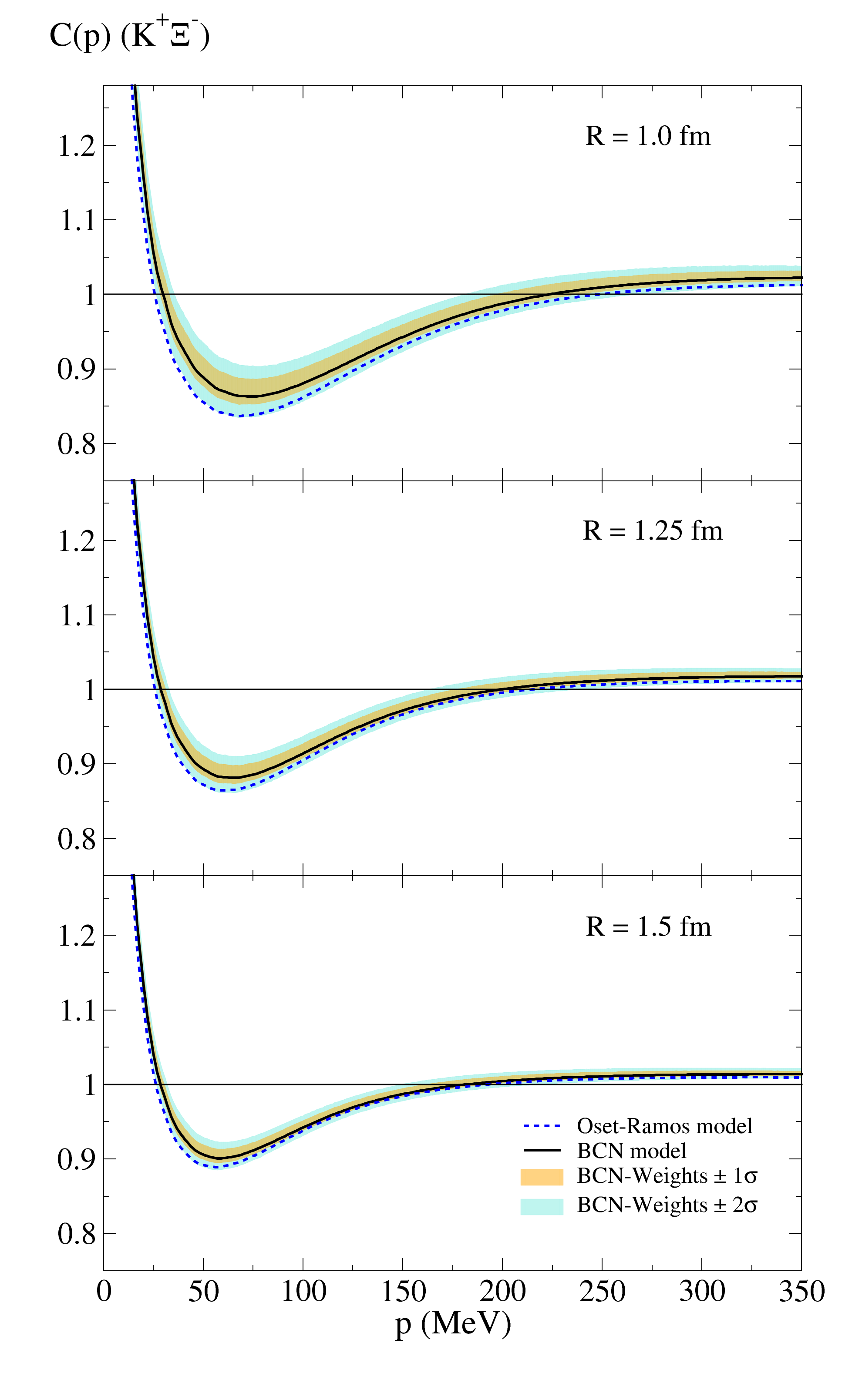}
    \caption{$K^+\Xi^-$ CFs for the BCN and Oset-Ramos models (black and blue lines respectively), as well as the error bands associated to BCN model (see details in the text), for three values of the source size $R$.}
    \label{fig:KX full CF}
\end{figure}

Lastly, we present our prediction for the coupled channel contributions to the $K^+\Xi^-$ CF in Fig.~\ref{fig:KX channels}. In addition to the elastic channel, we observe that there is no sizable inelastic contribution, all of them amounting to a mere 12\% despite the sizable values for some of the associated production weights (see third column of the right hand panel in Table~\ref{tab:production_weights}). From this figure, it can be concluded that the elastic amplitude is essentially governing the CF behavior. 
 
Fig.~\ref{fig:KX kernel} shows the contribution of the different pieces of the interaction kernel of the BCN model to the $K^+\Xi^-$ CF, in the absence of the Coulomb interaction. First, it is important to remember that the $K^+\Xi^-$ elastic transition is dominated by a strong attractive contact term, which can be seen from the orange line in the present figure. Moreover, although the combination of the WT and NLO terms apparently describes the full CF without the need of the Born ones, the cyan line shows the importance of these terms when they are combined with the WT contribution. 

In Fig. \ref{fig:KX full CF} the complete $K^+\Xi^-$ CF is shown, taking into account the strong contribution predicted by the BCN and Oset-Ramos models combined with the Coulomb interaction. The error bands correspond to the propagated uncertainties of the parameters of the BCN model and the production weights of Table~\ref{tab:production_weights}. Due to the lack of experimental information on the source associated to this channel, a channel-independent single Gaussian source is assumed. To show the dependence of the CF on the source size, three different radii are displayed within the scope of reasonable source sizes in p-p collisions. It can be seen that both models give a similar description, although the availability of precise experimental data might help in resolving the two models. At the moment, as already commented, these similarities of the CFs from the two models lie in the fact of the WT dominance not only in the elastic $K^+\Xi^-$ transition but also in the $K^+\Xi^-\to K^0\Xi^0$ one. Actually, this correlation function is currently under experimental analysis by ALICE, which opens the possibility of accessing energies where information obtained through scattering experiments is scarce.

\section{Conclusions}
\label{sec:conclusions}
We have carried out a detailed study of the correlation functions for three meson-baryon pairs in the $S=-1$ sector. In all cases, we employed coupled-channel UChPT based models as inputs to calculate the corresponding CFs. In particular, we focus on the BCN model that has been constrained to many different observables and showed a notable predictive power in this sector. This work has also been complemented not only by a comprehensive analysis of the role played by each channel in the corresponding CF but also by the relevance of the different contributions in the interaction kernel.

First, we revisited the theoretical analysis of the available experimental $K^-p$ CF existing in the literature. In contrast to the previous analysis, we showed that the BCN model is capable of describing the experimental data within 2$\sigma$ discrepancy, thereby dispelling any doubt about the reliability of the current chiral models to describe the meson-baryon interactions in this sector. As a matter of fact, our result is indirectly demonstrating the compatibility among the scattering cross-sections, the measurement of the $\bar{K}N$ threshold observables and the femtoscopic data. This fact gives us confidence in the potential of the femtoscopy technique to provide future constraints on theoretical scattering amplitudes, especially for those cases where scattering experiments are not feasible.

We have also provided, for the first time, predictions for the $K^+\Xi^-$ and the $\pi^-\Lambda$ CFs with the corresponding estimation of the error bands considering the uncertainties of the model parameters and the errors associated to the production weights. Such observables are currently under analysis by the ALICE collaboration, whose future comparison with the theoretical predictions will certainly shed some light on the almost uncharted meson-baryon interaction below the $\bar{K}N$ threshold and provide valuable insights at higher energies, where the available $K^-p \to K\Xi$ cross section data have large uncertainties. 

Additionally, we have presented two approaches, referred to as the VM and VLC methods, that deliver the amount of produced hadron pairs relative to the measured one. These production weights are essential ingredients in the theoretical computation of the correlation functions. The methods developed in the present study show to be compatible, although the VLC one acquires a more general character by construction.

We are just at the beginning of the LHC RUN3 data taking in a higher precision era and it is most probable that exciting outputs will be obtained in the near future. The interpretation of the experiments and the subsequent studies to learn about the nature of the $\bar{K}N$ interaction is a task that will require the combined efforts of both experimentalists and theoreticians, to which the present work aims at contributing.

\begin{acknowledgments}

P. E. is supported by the Spanish Ministerio de Ciencia e Innovaci\'on (MICINN) and European FEDER funds under Contracts No.\,PID2020-112777GB-I00, PID2023-147458NB-C21 and CEX2023-001292-S (Unidad de Excelencia “Severo  Ochoa”); by Generalitat Valenciana under contract CIPROM/2023/59. A. F. was supported by ORIGINS cluster DFG under Germany’s Excellence Strategy-EXC2094 - 390783311 and the DFG through the Grant SFB 1258 ``Neutrinos and Dark Matter in Astro and Particle Physics”. V. M. S. was supported by the Deutsche Forschungsgemeinschaft (DFG) through the grant MA $8660/1-1$. A. R. acknowledges support from MICIU/AEI/10.13039/501100011033 and by FEDER UE through grant PID2023-147112NB-C21 and through the ``Unit of Excellence Mar\'ia de Maeztu 2020-2023" award to the Institute of Cosmos Sciences, grant CEX2019-000918-M. 

\end{acknowledgments}

\appendix

\section{Production weights}
\label{app:weights}

For a given observed channel $i$, the $j$-th production weight $w_j$ accounts for the number of $j$ pairs that are produced during the hadronization phase, such that they are susceptible to interact, to transform into the observed pair and contribute to the final CF \cite{Fabbietti:2020bfg}. These weights, in general, can be split into two components,
\begin{equation}
    w_j = w_j^{T}\times w_j^{kin}\ .
\end{equation}
Here, $w_j^{T}$ accounts for the number of $j$ pairs  (relative to $i$ pairs) that are produced in the collision, and $w_j^{kin}$ accounts for the fraction of pairs that are kinematically available to interact and transform into the final pair. Note that the transition probability from one pair into another does not have to be accounted for in the weights, since it is implicit in the scattering amplitude in the KP formula.

The thermal production of particles in p-p collisions is estimated within the statistical hadronization model included in the Thermal-FIST framework \cite{Vovchenko:2019pjl}. Following \cite{Vovchenko:2019kes}, in smaller colliding systems, a good description of the particle production yields is achieved within the canonical ensemble assuming an incomplete chemical equilibrium of strangeness ($\gamma_S$CSM model). This model depends on three parameters: the chemical freeze-out temperature $T_{ch}$, the strangeness saturation parameter $\gamma_S$ and the volume parameter at midrapidity $dV/dy$. In \cite{Vovchenko:2019kes}, a parametrization of these quantities is given in terms of the average multiplicity of charged particles, $\langle dN_{ch}/d\eta\rangle_{|\eta|<0.5}$. The experimental $K^-p$ CF was measured with a minimum-bias trigger, leading to $\langle dN_{ch}/d\eta\rangle=6.91$ \cite{ALICE:2022yyh}, while more recent ALICE measurements, such as the preliminary ones of the $K^+\Xi^-$~\cite{SQMXiPiALICE}, are selected with a trigger for high-multiplicity events, giving an average production of charged particles of $\langle dN_{ch}/d\eta\rangle\simeq30$ (see Ref.~\cite{ALICE:2020mfd}). The corresponding $\gamma_S$CSM parameters are displayed in Table~\ref{tab:parameters}.

This way, the thermal component of the weights is computed as
\begin{equation}
    w_j^{T} = \frac{N_j^T}{N_i^T}
\end{equation}
where $N_j^{T}$ is the product of the primary production yields of the two particles forming the $j$-th pair.

Analogously to the previous equation, the kinematic component of the weights can be written as
\begin{equation}
    w_j^{kin} = \frac{N_j^{kin}}{N_i^{kin}}.
\end{equation}
Given the normalized relative momentum distribution of the $j$-th pair, $dN_j/dk^{*} _j$, then $N_j$ is the integral of this distribution over the relative momentum range of the $i$-th pair covered by the observed CF. That is,
\begin{equation}
    N_j = \int_{0}^{k^*_{max}} dk_i^* \frac{dN_j}{dk_i^*}\frac{dk_i^*}{dk_j^*}.
\end{equation}
Typically, the signal due to the final-state interaction of the CF is visible at low momenta, and converges to unity around $k^*_i\sim 300$ MeV/c. For this reason, we integrate the distributions up to $k^*_{max}=300$ MeV/c.

The $k^*$ distributions are obtained from the individual momenta of the particles forming the pair, $\vec{p}=(p_T\cos\phi,\ p_T\sin\phi,\ m_T\sinh y)$, where $p_T$ is the transverse momentum, $m_T=(p_T^2+m^2)^{1/2}$ and $y$ is the rapidity. Following \cite{Schnedermann:1993ws}, a stationary thermal model is unable to reproduce the experimental production of hadrons, since the direction of the collision is embedded in the hadronization process. Instead, a longitudinal expansion model, often referred to as the cylindrically symmetric blast-wave (CBW) model, fits the data best. Within this model, the three-momentum spectrum is given by a uniform azimuthal angle and the following distribution
\begin{eqnarray}
    &\displaystyle\frac{d^2N}{dp_Tdy}& \propto p_T m_T \int_0^1 \tilde{r}d\tilde{r}\int_{-\tilde{\eta}_{max}}^{\tilde{\eta}_{max}} d\tilde{\eta}\cosh(\tilde{\eta}-y) \nonumber \\
    && \times \exp\left[ -\frac{m_T\cosh\rho\cosh(\tilde{\eta}-y)}{T_{kin}} \right] I_0\left(\frac{p_T\sinh\rho}{T_{kin}}\right)\ , \nonumber \\
    \label{eq:fullCBW}
\end{eqnarray}
where $I_0(x)$ is the modified Bessel function of the first kind and $\rho=\tanh^{-1}(\beta_S\tilde{r}^n)$. Here, $\beta_S\tilde{r}^n$ is the fireball transverse flow velocity profile. This distribution depends on four free parameters: the velocity profile exponent $n$; the kinetic freeze-out temperature $T_{kin}$; the transverse flow velocity at the surface $\beta_S$, which is often parametrized through the mean transverse flow velocity, $\langle\beta_T\rangle=\frac{2}{2+n}\beta_S$; and the longitudinal rapidity cutoff $\tilde{\eta}_{max}$.

\begin{table}[h]
\begin{center}
\begin{tabular}{|c|c|c||c|c|}
\hline
Channel & \hspace{0.2cm} $K^-p$ CF \hspace{0.2cm} & \hspace{0.2cm} $K^+\Xi^-$ CF \hspace{0.2cm} &  Channel & \hspace{0.2cm} $\pi^-\Lambda$ CF \hspace{0.2cm} \\ [2mm] \hline
 $K^-p$  & $1$ & $6.35^{+1.92} _{-1.35}$ &$\pi^-\Lambda$& $1$\\  [1mm]
 $\bar{K}^0n$& $0.935^{+0.02} _{-0.02}$ & $6.28^{+1.84} _{-1.33}$ &$\pi^0\Sigma^-$& $0.45^{+0.09} _{-0.06}$\\  [1mm]
$\pi^0\Lambda$& $1.40^{+0.18} _{-0.15}$ & $6.91^{+2.93} _{-1.84}$ &$\pi^-\Sigma^0$& $0.46^{+0.01} _{-0.01}$\\ [1mm]
$\pi^0\Sigma^0$& $1.06^{+0.1} _{-0.06}$ & $5.43^{+1.99} _{-1.31}$ &$K^-n$& $0.143^{+0.012} _{-0.010}$\\  [1mm]
 $\pi^+\Sigma^-$& $0.95^{+0.08} _{-0.07}$ & $5.21^{+1.89} _{-1.27}$ &$\eta\Sigma^-$& $0$\\  [1mm]
$\pi^-\Sigma^+$& $0.98^{+0.09} _{-0.07}$ & $5.34^{+1.97} _{-2.51}$ &$K^0\Xi^-$& $0$\\ [1mm]
 $\eta\Lambda$  & $0$ & $2.54^{+0.29} _{-0.22}$ &$\_$ & $\_$ \\ [1mm]
$\eta\Sigma^0$& $0$ & $1.88^{+0.12} _{-0.08}$ &$\_$ & $\_$\\ [1mm]
$K^+\Xi^-$& $0$ & $1$ &$\_$ & $\_$\\ [1mm]
$K^0\Xi^0$& $0$ & $1.03^{+0.01}_{-0.01}$ &$\_$ & $\_$ \\  [1mm]
\hline
\end{tabular}
\caption{Central values of the production weights for $\pi^-\Lambda$, $K^-p$ and $K^+\Xi^-$ CFs following the VM Method.}
\label{tab:weights_methodVM}
\end{center}
\end{table} 

\begin{table*}[]
    \centering
    \begin{tabular}{|c|c|c|c|} \cline{3-4}
        \multicolumn{1}{c}{} & & Minimum bias ($K^-p$) & High multiplicity ($K^+\Xi^-$ and $\pi^-\Lambda$) \\ \hline
        \multirow{3}{*}{$\gamma_S$CSM} & $T_{ch}$(GeV) & $0.171\pm0.001$ & $0.167\pm0.001$ \\
        & $\gamma_S$ & $0.78\pm0.01$ & $0.85\pm0.01$ \\
        & $dV/dy$ (fm$^3$) & $17\pm1$ & $72\pm6$ \\ \hline
        \multirow{4}{*}{CBW (VM Method)} & $T_{kin}$ (GeV) & $0.166^{+0.02} _{-0.01}$ & $0.151^{+0.021} _{-0.011}$ \\
        & $\beta_S$ & $0.767\pm0.049$ & $0.828\pm0.072$ \\
        & $n$ & $3.9^{+0.32} _{-0.76}$ & $1.24\pm0.08$ \\
        & $\tilde{\eta}_{max}$ &  $\to\infty$ & $\to\infty$  \\ \hline
        \multirow{4}{*}{CBW (VLC Method)} & $T_{kin}$ (GeV) & $0.136\pm0.003$ & $0.168\pm0.004$ \\
        & $\beta_S$ & $0.895\pm0.011$ & $0.836\pm0.019$ \\
        & $n$ & $4.85\pm0.23$ & $2.86\pm0.29$ \\
        & $\tilde{\eta}_{max}$ & $1.7$ & $1.7$ \\ \hline
    \end{tabular}
    \caption{Parameters of the $\gamma_S$CSM and CBW models for minimum-bias and high-multiplicity trigger events. Values in the VM method are taken from~\cite{ALICE:2013wgn}. The total error is evaluated as the sum in quadrature of the reported statistical and systematic uncertainties. See the text for details.}
    \label{tab:parameters}
\end{table*}

\begin{figure}
    \centering
    \includegraphics[width=\columnwidth]{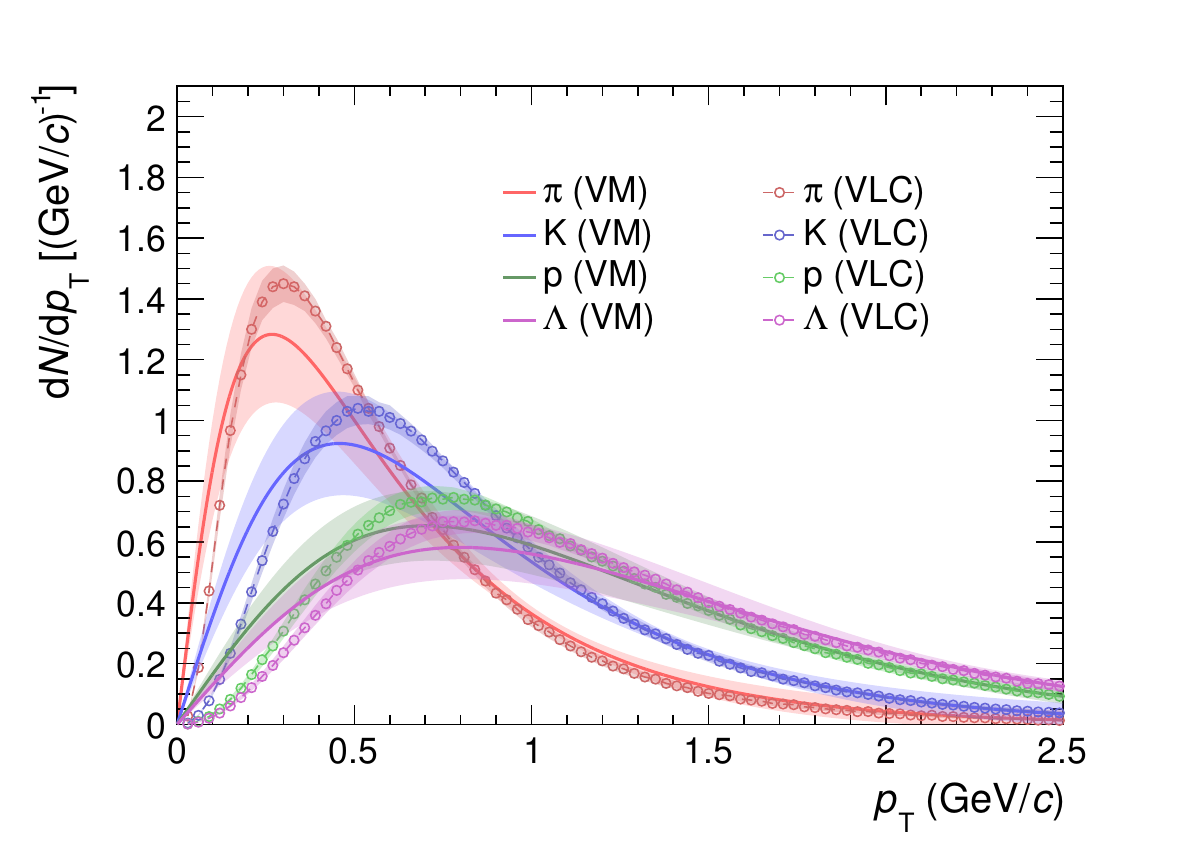}
    \caption{Fitted $p_T$ distribution in VM and VLC methods for four particle species: $\pi$, $K$, $p$ and $\Lambda$. The HM parametrization for the BW and TF models is used.}
    \label{fig:pTfit}
\end{figure}

\begin{figure*}
    \centering
    \includegraphics[width=\columnwidth]{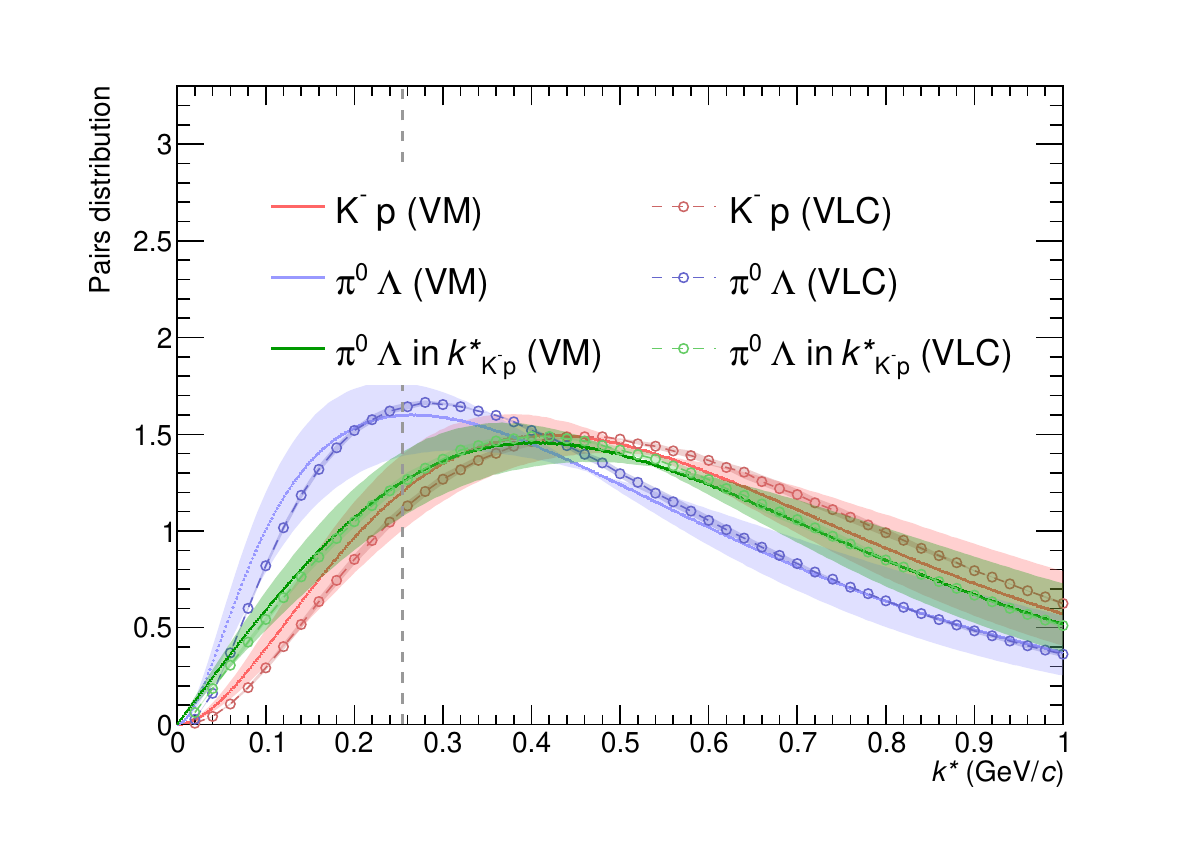}
    \includegraphics[width=\columnwidth]{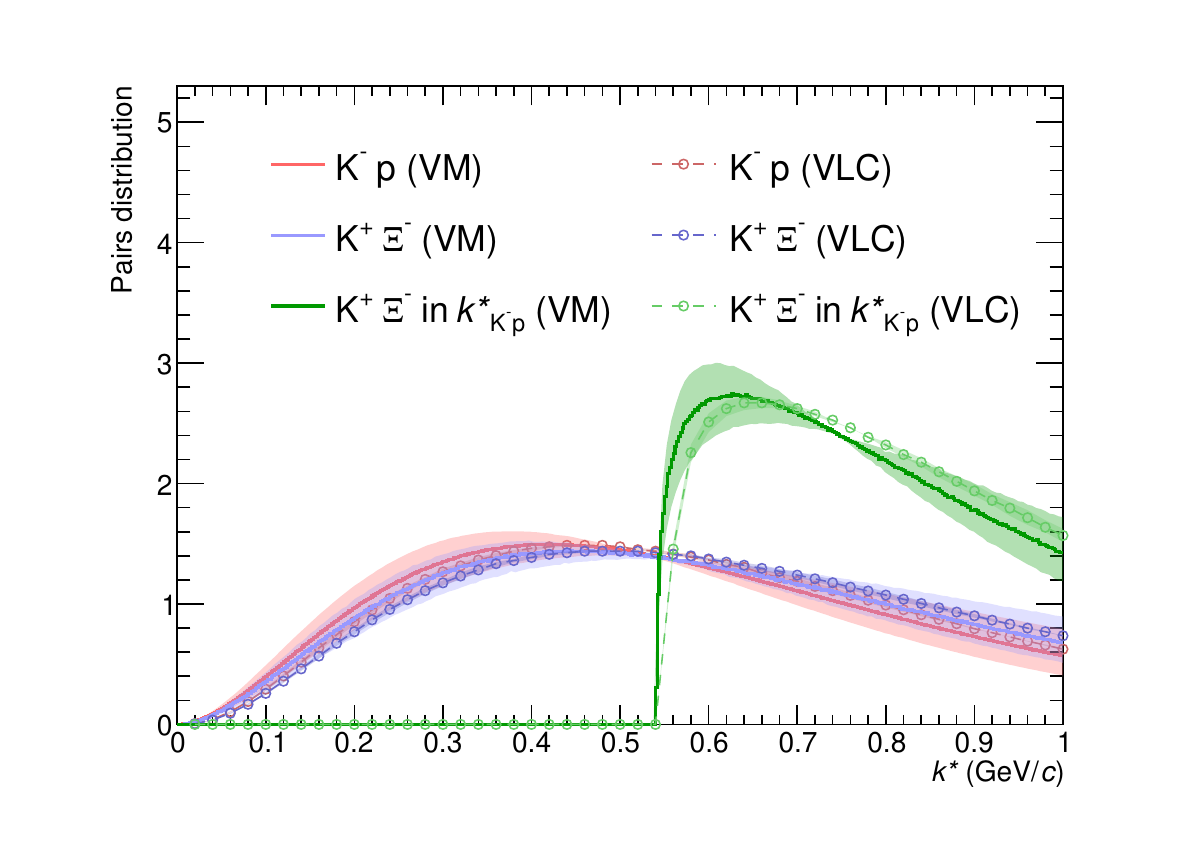}
    \caption{Distribution in $k^\ast$ for the $K^-p$ pairs in comparison to the lowest $\pi^0\Lambda$ and $K^+\Xi^-$ highest energy channels. Left: comparison to the $\pi^0\Lambda$ with the grey vertical line indicating the minimum $k^{\ast} _{\pi^0\Lambda}= 0.254$ GeV/c. Right: comparison to the $K^+\Xi^-$. The bands are obtained from varying within uncertainties the parameters of the BW and TF models. The HM parametrization for the BW and TF models is used.}
    \label{fig:kStarDistrib}
\end{figure*}

We have followed two approaches, which turn out to be compatible and deliver production weights in agreement within the corresponding uncertainties. The reason to employ two different way to evaluate the weights is to provide robust methods which can be used both, either when input from experimental spectra measurements are available (e.g $p_T$ spectra) or when a parametrization of BW fits is provided directly.

The first one, referred to as the VM Method (Valencia-Munich), assumes a boost-invariant CBW model, obtained from taking the limit $\tilde{\eta}_{max}\to\infty$, so that the previous distribution reduces to
\begin{equation}
    \frac{dN}{dp_T} \propto p_T m_T \int_0^1  I_0\left(\frac{p_T\sinh\rho}{T_{kin}}\right) K_1\left(\frac{m_T\cosh\rho}{T_{kin}}\right)\tilde{r}d\tilde{r}\ ,
\end{equation}
and a uniform pseudorapidity distribution, $\eta\in[-0.5,0.5]$, corresponding to the used interval of pseudorapidity for spectra measurements~\cite{ALICE:2013wgn}. Additional checks on extending this range to the ALICE acceptance $|\eta|<0.8$ deliver a negligible modification on the weights, still within the quoted uncertainties in Tab.~\ref{tab:weights_methodVM}.  Note that, since the pseudorapidity $\eta$ is connected to the rapidity $y$, this cut limits the $z$-component of the three-momentum\footnote{In this case the single momenta of the particles in the pair read $\vec{p}=(p_T\cos\phi,\ p_T\sin\phi,\ p_T\sinh \eta)$}. Here $K_1(x)$ is the modified Bessel function of the second kind. The remaining free parameters have been fitted in \cite{ALICE:2013wgn} to experimental $p_T$ spectra in p-Pb collisions. We assume p-p collisions to be equivalent to the event class in p-Pb that has the same average multiplicity of charged particles as the studied p-p events. That is, to compute the $K^-p$ weights we choose the 80-100\% event class (minimum-bias), while for the $K^+\Xi^-$ and $\pi^-\Lambda$ weights we choose the 10-20\% event class (high-multiplicity). These parameters are shown in Table~\ref{tab:parameters} (VM Method), and the values of the weights obtained through the VM Method are shown in Table~\ref{tab:weights_methodVM}.

For the second approach, referred to as the VLC Method (Valencia), we consider the full CBW distribution of Eq.~(\ref{eq:fullCBW}) and generate events keeping only those satisfying $|\eta|<0.5$. This cut implicitly changes the functional form of the $p_T$ distribution and thus the set of parameters used in the previous method are no longer valid, since we would no longer reproduce the experimental spectra. Besides, now $\tilde{\eta}_{max}$ is finite, breaking the boost invariance, and is fixed to $1.7$, a value fitted to experimental rapidity distributions for different particle species \cite{Schnedermann:1993ws}, obtaining a result independent of the temperature.

In order to be consistent with the experimental spectra given in~\cite{ALICE:2013wgn}, we fit the resulting $p_T$ BW distribution to synthetic data extracted from the $p_T$ distributions of the first method, see Fig.~\ref{fig:pTfit}. As can be seen, the VM and VLC spectra show an overall agreement, with only a slight tension in the very low $p_T$ region.
After applying the $\eta$ cut, the VLC $p_T$ spectrum reproduces the experimental one while maintaining the intrinsic angular dependence of the CBW model. The fitted parameters are shown in Table~\ref{tab:parameters} (VLC Method). It is worth mentioning that these parameters are not to be interpreted physically, as their only function is compensating the hardening of the $p_T$ spectrum. The final values of the production weights, obtained with VLC Method, are displayed in Table~\ref{tab:production_weights}. These values can be compared with those of Table~\ref{tab:weights_methodVM} to see the compatibility between both methods.

For completeness, we also report in Fig.~\ref{fig:kStarDistrib} the resulting VM and VLC $k^\ast$ distributions of $K^-p$ pairs (red bands and markers) in comparison to the distribution of the lowest channel $\pi^0\Lambda$ (left) and of the highest channel $K^+\Xi^-$ (right). The distribution of both $\pi^0\Lambda$ and $K^+\Xi^-$ systems as a function of their own relative momentum is presented in blue. The green bands indicates the resulting distributions of each of these channels transformed into the final measured $K^-p$ pair, which is one of the ingredients needed to evaluate the production weights.
It can be appreciated the compatibility between all these distributions in both models which confirm the agreement achieved at the level of the weights' calculations.

\bibliography{apssamp}

\end{document}